\documentclass{aastex631} 
\usepackage{amsmath}
\usepackage{multirow,tabularx}
\usepackage{hyperref}

\begin{document}

\title{Doomed Worlds I: No new evidence for orbital decay in a long-term survey of 43 ultra-hot Jupiters}

\correspondingauthor{Elisabeth Adams}
\email{adams@psi.edu}

\author[0000-0002-9131-5969]{Elisabeth R. Adams}
\affiliation{Planetary Science Institute\\ 
1700 E. Ft. Lowell, Suite 106, \\
Tucson, AZ 85719, USA}

\author[0000-0002-9495-9700]{Brian Jackson}
\affiliation{Department of Physics, Boise State University\\ 
1910 University Drive, \\ 
Boise ID 83725-1570 USA}

\author[0000-0002-9468-7477]{Amanda A. Sickafoose}
\affiliation{Planetary Science Institute\\ 
1700 E. Ft. Lowell, Suite 106, \\
Tucson, AZ 85719, USA}

\author[0000-0003-3716-3455]{Jeffrey P. Morgenthaler}
\affiliation{Planetary Science Institute\\ 
1700 E. Ft. Lowell, Suite 106, \\
Tucson, AZ 85719, USA}

\author{Hannah Worters}
\affiliation{South African Astronomical Observatory\\ 
1 Observatory Rd., Observatory, 7945 South Africa}

\author[0009-0005-2475-4987]{Hailey Stubbers}
\affiliation{Department of Physics, Boise State University\\ 
1910 University Drive, \\ 
Boise ID 83725-1570 USA}

\author[0009-0001-5881-1856]{Dallon Carlson}
\affiliation{Department of Physics, Boise State University\\ 
1910 University Drive, \\ 
Boise ID 83725-1570 USA}

\author[0000-0002-6673-8206]{Sakhee Bhure}
\affiliation{University of Southern Queensland, Centre for Astrophysics, West Street, Toowoomba, QLD 4350 Australia}

\author{Stijn Dekeyser}
\affil{University of Southern Queensland, Centre for Astrophysics, West Street, Toowoomba, QLD 4350 Australia} 

\author[0000-0003-0918-7484]{Chelsea X. Huang}
\affiliation{University of Southern Queensland, Centre for Astrophysics, West Street, Toowoomba, QLD 4350 Australia}

\author[0000-0001-9194-2084]{Nevin N. Weinberg}
\affiliation{Department of Physics, University of Texas at Arlington, Arlington, TX 76019, USA}

\begin{abstract}
Ultrahot Jupiters (UHJs) are likely doomed by tidal forces to undergo orbital decay and eventual disruption by their stars, but the timescale over which this process unfolds is unknown. We present results from a long-term project to monitor UHJ transits. We recovered WASP-12 b's orbital decay rate of $\dot P = -29.8\pm1.6$ ms yr$^{-1}$, in agreement with prior work. Five other systems initially had promising nonlinear transit ephemerides. However, a closer examination of two -- WASP-19 b and CoRoT-2 b, both with prior tentative detections -- revealed several independent errors with the literature timing data; after correction neither planet shows signs of orbital decay. Meanwhile, a potential decreasing period for TrES-1 b, $\dot P = -16\pm5$ ms yr$^{-1}$, corresponds to a tidal quality factor $Q_\star'=160$ and likely does not result from orbital decay, if driven by dissipation within the host star. Nominal period increases in two systems, WASP-121 b and WASP-46 b, rest on a small handful of points. Only 1/43 planets (WASP-12 b) in our sample is experiencing detectable orbital decay. For nearly half (20/42) we can rule out $\dot P$ as high as observed for WASP-12 b. Thus while many ultra-hot Jupiters could still be experiencing rapid decay that we cannot yet detect, a sizable sub-population of UHJs are decaying at least an order of magnitude more slowly than WASP-12 b. Our reanalysis of Kepler-1658 b with no new data finds that it remains a promising orbital decay candidate. Finally, we recommend that the scientific community take steps to avoid spurious detections through better management of the multi-decade-spanning datasets needed to search for and study planetary orbital decay.
\end{abstract}



\section{Introduction} \label{sec:Introduction}

Ultrahot Jupiters (UHJs), or giant planets with orbital periods less than about 3 days, experience significant tidal effects that play a critical role in their long-term dynamical stability. Recent work has shown that the population of all hot Jupiter ($P<10$ d) host stars is younger than the general population of either field stars or planet-hosting stars \citep{2019AJ....158..190H}, and two separate analyses have found lower UHJ occurrence rates around older host stars \citep{2023PNAS..12004179C, 2023AJ....166..209M}. All of this supports the idea that many giant, close-in planets quickly inspiral or are otherwise destroyed while their stars are still on the main sequence. Meanwhile, an estimated half of all stars may have ingested a former UHJ in the first $<1$ Gyr of the star's lifetime \citep{2015ApJ...809L..20M}. Recent direct detection of a planetary engulfment by \citet{De2023} for an inferred planet of 0.1-10 $M_J$ came in the form of a low-luminosity optical transient lasting several days, followed by infrared brightening for several months. Such planet engulfment events are thought to occur somewhere in the galaxy about once every 1-10 years \citep{2012MNRAS.425.2778M, De2023}. Estimates for the lifetimes of UHJs as a population have been limited, however, by theoretical uncertainties in the value of the stellar tidal dissipation factor, $Q_\star^\prime$, where estimates range from $10^{5.5}-10^{6.5}$ \citep{2008ApJ...678.1396J, 2012MNRAS.422.3151H} to $>10^7-10^8$ \citep{2012ApJ...751...96P, 2018MNRAS.476.2542C}. More recent theoretical work that incorporates nonlinear dissipation has found important variations in $Q_\star^\prime$ with stellar mass and age and that planets around stars with $M_\star \ge 1.2 M_{Sun}$ only experience significant tidal decay when the star is on the subgiant branch \citep{2024ApJ...960...50W}. \citet{2018AJ....155..165P} found a steep dependence on forcing frequency for $Q_\star^\prime$, with $\log_{10} Q_\star^\prime \sim 5$ for orbital periods $P \approx 2\,{\rm days}$ and increasing rapidly to $\log_{10} Q_\star^\prime \sim 7.5$ for $P \approx 0.5\,{\rm day}$. Consequently, tidal decay might be expected to slow as the planet nears its star. It is important to note, however, that for planets whose rotation states are not tidally locked or whose orbits are eccentric, tidal decay can be driven by dissipation within the planet. However, tidal locking \citep{2002A&A...385..156G} and orbital circularization \citep{2018ARA&A..56..175D} timescales for hot Jupiters are thought to be less than millions of years. Thus, in the absence of an exotic rotation state \citep{2018ApJ...869L..15M, 2022Univ....8..211E} or dynamical excitation of eccentricity \citep{2019MNRAS.488.3568P}, the contribution to tidal decay from dissipation within the planets is likely to be short-lived.

With a known population of around 100 UHJs, it is now possible to empirically address the open questions of how many UHJs are in decaying orbits and at what rates they are decaying. Strong evidence for decay has been presented for just two UHJs, WASP-12 b \citep{2017AJ....154....4P, 2019MNRAS.490.1294B, 2020ApJ...888L...5Y} and Kepler-1658 b \citep{Vissapragada_2022}, with orbital periods that are apparently decreasing by $-29\pm2$ ms yr$^{-1}$ and $-131\pm22$ ms yr$^{-1}$, respectively, corresponding to remaining lifetimes of 3 Myr and 2.5 Myr. These decay rates also correspond to stellar tidal dissipation parameters of $Q_\star^\prime = 1.8\times10^5$ and $2.5\times10^4$, respectively. Kepler-1658 b orbits an evolved subgiant ($M_\star=1.45~ M_{Sun}$, $R_\star = 2.89~ R_{Sun}$), for which that rapid rate of decay matches the theoretical predictions of \citet{Vissapragada_2022}, though perhaps not those of \citet{2024MNRAS.527.5131B} (see discussion in Section \ref{sec:kepler1658b}). Few planets are known around recently evolved stars due to challenges with their detection \citep[see e.g.][]{2007ApJ...665..785J}. In fact, one explanation for the evolution of WASP-12 b may be that its host star is also a subgiant \citep{2017ApJ...849L..11W}, though modeling and observational uncertainties leave its status ambiguous \citep{2024A&A...686A..84L}. This raises an open question: are there any ultra-hot Jupiters around main-sequence stars with decaying orbits?

In the past few years, as observational baselines have passed the decade mark for many systems, a growing number of UHJs around main-sequence stars have been presented with suggestions that they may have decreasing orbital periods. These include:  HAT-P-19 b \citep{2022AJ....164..220H}; HAT-P-32 b \citep{2022AJ....164..220H};  HAT-P-51 b \citep{2024NewA..10602130Y}; HAT-P-53 b \citep{2024NewA..10602130Y}; KELT-9 b \citep{2023A&A...669A.124H}; TrES-1 b \citep{2022AJ....164..220H, 2022ApJS..259...62I}; TrES-2 b \citep{2022AJ....164..220H};  TrES-3 b \citep[but noted to be marginal,][]{2022AJ....164..198M, 2022AJ....164..220H}; TrES-5 b \citep[nonlinearity though not decay noted by][]{2021A&A...656A..88M,  2022AJ....164..220H, 2022ApJS..259...62I, 2024NewA..10602130Y}; WASP-4 b \citep{2020ApJ...893L..29B,  2022AJ....164..220H, 2023A&A...669A.124H}; WASP-19 b \citep{2020AJ....159..150P, 2022ApJS..259...62I}; WASP-32 b \citep[but not significant,][]{2023MNRAS.520.1642S}; WASP-43 b as reported by \citet{2018ChA&A..42..101S} though not by \citet{2022AJ....164..220H}; and XO-3 b \citep{Yang_2022, 2022ApJS..259...62I}. Many of these claimed detections are acknowledged in the original publications to be of marginal significance, however. It is also common for an apparent non-linear ephemeris based on a small numbers of transits and/or a short baseline to disappear with additional observation \citep[e.g., as for OGLE-TR-113b,][]{2010ApJ...721.1829A,2016MNRAS.455.1334H}. Even after an ostensible period decay is detected, it is necessary to rule out other physical mechanisms, such as apsidal precession \citep[the current best explanation for KELT-9 b,][]{2023A&A...669A.124H}, perturbations caused by companion planets or stars, or even the acceleration of the host star towards the Earth \citep[proposed as the explanation for WASP-4 b,][]{2020ApJ...893L..29B}.

\begin{figure}
    \centering
    \includegraphics[width=\textwidth]{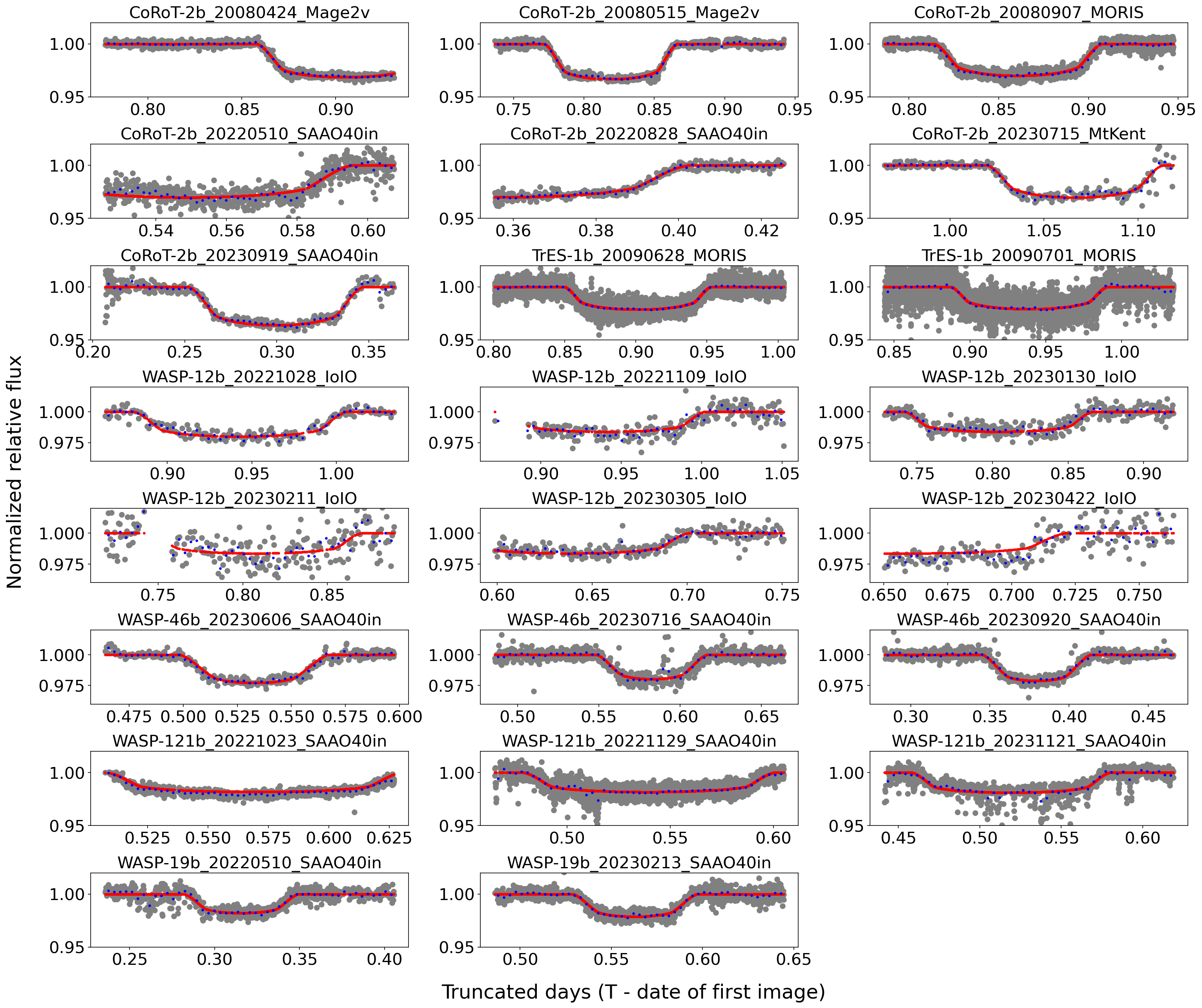}
    \caption{Twenty-three new light curves of six planets: CoRoT-2 b, TrES-1b, WASP-12 b, WASP-19 b, WASP-46 b, and WASP-121 b. The date and site of each observation are given in the panel header (see Section \ref{sec:Observations}). Normalized relative flux is plotted vs. truncated time, where the date of the first observation in each time sequence has been subtracted for clarity of display.  Data are shown as gray points, with binned data shown as blue points. A transit light curve model fit is overplotted in red. These light curves are available as supplementary data.} 
    \label{fig:all_transits}
\end{figure}

Detecting orbital decay rewards a patient approach. Recent theoretical work by \citet{2023AJ....166..142J} has explored how repeated, regular observations of transits over a long time scale will detect planets with relatively fast, WASP-12 b-like rates of orbital decay, requiring roughly two transits per year for 10+ years to build up clear evidence. Building on previous suggestions \citep{2007MNRAS.377L..74L}, that work focused on the Bayesian Information Criterion, or BIC \citep{1978AnSta...6..461S}, both to judge whether a dataset supports tidal decay and to design an observing program to optimally detect said decay. In this context, the BIC value can be defined as:

\begin{equation}
    {\rm BIC} = \chi^2 + k \ln N, \label{eqn:BIC}
\end{equation}

\noindent
where $N$ is the total number of data points, $k$ is the number of fit parameters (2 for a linear fit and 3 for a quadratic fit), and $\chi^2$ measures the goodness of fit. The BIC value thus favors models that minimize residuals while penalizing additional model parameters. In the analysis that follows, we use two models: one with a linear term in orbital epoch (a linear ephemeris), and one with a quadratic term in orbital epoch (a quadratic ephemeris). The equation for a linear ephemeris is

\begin{equation}
    T(E) = T_{0,lin} + P_{lin} \times E, \label{eqn:linear}
\end{equation}

\noindent
where the transit midtime, $T$, at a given epoch, $E$, may be predicted using the linear ephemeris reference time, $T_{0,lin}$, at $E=0$, and the orbital period, $P_{lin}$. The equation for a quadratic ephemeris is

\begin{equation}
    T(E) = T_{0,quad} + P_{quad} \times E + \frac{1}{2} \frac{dP}{dE} E^2,
    \label{eqn:quad}
\end{equation}

\noindent
where $dP/dE$ is the rate of change of the orbital period, $P_{quad}$, and $T_{0,quad}$ is the quadratic ephemeris reference time.

Note that while the parameter used in the fit is typically $dP/dE$ (which has units of days per epoch), the value most frequently quoted in the literature is the rate of change in ms yr$^{-1}$, or 



\begin{equation}
   \dot P = \left( \frac{dP}{dE} \right)[{\rm d\ \rm{epoch}^{-1}}] \left( \frac{365.25}{P} \right)[{\rm epoch\ \rm yr^{-1}}] (86400) [{\rm s\ d^{-1}}] (1000) [{\rm ms\ s^{-1}}]
    \label{eqn:pdot}
\end{equation}

where $P$ is the orbital period in days, $dP/dE$ is in days per epoch, and $\dot P$ is the rate of change of $P$ in ms yr$^{-1}$. 

To judge whether a linear or a quadratic ephemeris is favored by the data, we use the difference in the BIC values between the linear model (${\rm BIC}_{\rm lin}$) and the quadratic model (${\rm BIC}_{\rm quad}$):

\begin{equation}
    \Delta {\rm BIC} =  {\rm BIC}_{\rm lin} - {\rm BIC}_{\rm quad}. \label{eqn:deltaBIC}
\end{equation}

\noindent
A positive value of $\Delta {\rm BIC}$ means that a quadratic ephemeris is preferred, and the higher the $\Delta {\rm BIC}$ value, the higher the probability that the quadratic model (indicating possible orbital decay) is supported by the data over the linear (no tidal decay) model.

In this work, we present the first results of a new, long-term program to take regular ground-based observations of UHJs and combine these data with times from the literature, which for some planets stretch back over two decades. We have observed more thana hundred new light curves of 43 UHJs for this project since 2022, and we are also publishing a cache of older light curves from 2008-2009 that had previously eluded publication. For this first paper, we summarize the results for all 43 systems but leave an in-depth analysis of most systems for future work. We focus here on six planets -- CoRoT-2 b, TrES-1 b, WASP-12 b, WASP-19 b, WASP-46 b, and WASP-121 b --  all of which initially had values for $\Delta {\rm BIC}\ge30$ from calculations using our new midtimes and the midtimes recorded in the literature. 

It is important to state from the outset that for two of the six planets we chose to focus on in this work -- CoRoT-2 b and WASP-19 b, which initially had the largest $\Delta {\rm BIC}$ values after WASP-12 b -- we needed to correct numerous errors and inconsistencies in the literature. After doing so, our final $\Delta {\rm BIC}$ values for both planets dropped so significantly  that neither one of them now shows any signs of nonlinear ephemerides.  WASP-12 b remains the \emph{only} candidate in our sample with evidence for orbital decay. (Notably, our sample does not include Kepler-1658 b, for which we have no new observations: its depth of 0.1\% makes it difficult to observe except from space or from the largest ground-based observatories. We do however apply the same analysis to the existing data for Kepler-1658 b and find it to be a very promising candidate, as discussed in Section \ref{sec:kepler1658b}.)

In Section~\ref{sec:Observations}, we discuss the new ground-based observations and describe the photometry methods used to create the new transit light curves. Data from TESS and CoRoT are described in Section~\ref{sec:space}. We describe our transit light curve fitting method in Section~\ref{sec:fit}. In Section~\ref{sec:lit-data}, we describe challenges in compiling the literature data and detail several issues that we uncovered with published transit midtime values. In Section~\ref{sec:results}, we describe our best current analysis of the timing for each of the six planets we are exploring in detail, plus Kepler-1658 b. In Section~\ref{sec:population_and_qstar}, we discuss what the results mean for the population of ultra-hot Jupiters and the values of $Q_\star'$ for their host stars. As discussed above, tidal dissipation within the planet could contribute to decay, but we have no evidence either that the planets we study here are not tidally locked or have eccentric orbits, so we neglect this contribution. Consequently, the constraints we report on $Q^\prime_\star$ represent lower limits. We conclude in Section~\ref{sec:disc} with some recommendations for best practices for searches for orbital decay, and a proposal for better management of transit timing data to prevent unnecessary errors in the future, which will enable the detection of subtle timing effects that unfold over many years across large, inhomogeneous datasets.



\begin{table}
\caption{Observational Details}
\begin{tabular}{llrrrrrrrrr} 
\hline
Planet & Date$^a$ & Instr. & Time  & $N_{\rm frames}$ & $N_{\rm frames}$ & $N_{\rm c}$$^b$ & Filter & Exp.  & Aper. & Scatter$^c$  \\
 & & & (hr) &  (use) & (omit) & &  &  (s) & (px) & (\%)\\
\hline
CoRoT-2b & 2008-04-24 & Mage2v & 3.7 & 1156 & 0 & 1 & i & 5-10 & 18 & 0.19\\
CoRoT-2b & 2008-05-15 & Mage2v & 4.9 & 1214 & 0 & 1 & i & 4-10 & 16 & 0.2\\
CoRoT-2b & 2008-09-07 & MORIS & 3.9 & 2707 & 0 & 1 & R & 5 & 26 & 0.32\\
CoRoT-2b & 2022-05-11 & SAAO40in & 2.0 & 755 & 0 & 4 & R & 8-10 & 18 & 0.74\\
CoRoT-2b & 2022-08-28 & SAAO40in & 2.3 & 410 & 40 & 1 & R & 5-25 & 18 & 0.27\\
CoRoT-2b & 2023-07-15 & MINERVA & 3.7 & 203 & 0 & 2 & Clear & 60 & 21 & 0.47\\
CoRoT-2b & 2023-09-19 & SAAO40in & 3.8 & 357 & 0 & 1 & R & 8-60 & 8 & 0.55\\
TrES-1b & 2009-06-28 & MORIS & 4.9 & 9993 & 0 & 1 & R & 1 & 32 & 0.43\\
TrES-1b & 2009-07-01 & MORIS & 4.6 & 7994 & 0 & 1 & R & 2 & 36 & 0.82\\
WASP-12b & 2022-10-28 & IoIO & 4.3 & 179 & 3 & 8 & R & 42.71 & NA$^d$ & 0.32\\
WASP-12b & 2022-11-09 & IoIO & 4.4 & 276 & 5 & 1 & R & 42.71 & NA$^d$ & 0.55\\
WASP-12b & 2023-01-30 & IoIO & 4.9 & 329 & 0 & 12 & R & 42.71 & NA$^d$ & 0.4\\
WASP-12b & 2023-02-11 & IoIO & 4.1 & 221 & 28 & 1 & R & 42.71 & NA$^d$ & 1.11\\
WASP-12b & 2023-03-05 & IoIO & 3.6 & 237 & 2 & 8 & R & 42.71 & NA$^d$ & 0.41\\
WASP-12b & 2023-04-22 & IoIO & 3.6 & 202 & 14 & 16 & R & 42.71 & NA$^d$ & 0.82\\
WASP-46b & 2023-06-06 & SAAO40in & 3.2 & 357 & 0 & 1 & R & 20-45 & 8 & 0.26\\
WASP-46b & 2023-07-16 & SAAO40in & 4.3 & 761 & 0 & 1 & R & 15-35 & 8 & 0.43\\
WASP-46b & 2023-09-20 & SAAO40in & 4.4 & 615 & 0 & 1 & R & 20-35 & 12 & 0.32\\
WASP-121b & 2022-10-24 & SAAO40in & 2.9 & 1133 & 1 & 3 & R & 8-10 & 25 & 0.24\\
WASP-121b & 2022-11-29 & SAAO40in & 5.0 & 4062 & 210 & 1 & R & 3-6 & 15 & 0.48\\
WASP-121b & 2023-11-21 & SAAO40in & 4.3 & 1297 & 135 & 3 & R & 10-15 & 18 & 0.58\\
WASP-19b & 2022-05-10 & SAAO40in & 4.1 & 756 & 74 & 1 & R & 15-25 & 12 & 0.47\\
WASP-19b & 2023-02-13 & SAAO40in & 3.8 & 1316 & 10 & 3 & R & 8-15 & 10 & 0.4\\
\hline
\multicolumn{11}{l}{\footnotesize{a. Date of first observation in time series; may differ from transit name in \autoref{fig:all_transits} assigned by observing night.}} \\
\multicolumn{11}{l}{\footnotesize{b. Number of comparison stars used in light curve.}} \\ 
\multicolumn{11}{l}{\footnotesize{c. Standard deviation of residuals (normalized flux minus model).}} \\ 
\multicolumn{11}{l}{\footnotesize{d. Aperture photometry not used; see discussion in 
Section~\ref{sec:ioio}.}} 
\end{tabular}
\label{table:obs_settings}
\end{table}

\section{New Ground-based Observations and Photometry}
\label{sec:Observations}

In this paper, we are publishing the photometry for twenty-three new transits of six planets: CoRoT-2 b, TrES-1 b, WASP-12 b, WASP-19 b, WASP-46 b, and WASP-121 b, shown in \autoref{fig:all_transits}. (To allow for more in-depth analyses of individual systems, transit light curve photometry and full timing analyses for the other 37 planets in our sample will be published in future works. The results presented in this work for those 37 systems are based on at least one new light curve along with the literature transit midtimes compiled as in Section \ref{sec:lit-data}, although no attempt has been made to correct literature times except for removing obvious duplicates.) Five ground-based observatories produced these observations in 2008-2009 and 2022-2023, with observational details shown in \autoref{table:obs_settings}. All ground-based, TESS, and CoRoT light curves for the six planets featured in this work are available as supplementary data, with a stub table for format in \autoref{table:lc_stub}.

\begin{table}
\caption{Transit light curves used in fits in this work for six planets$^{a,b}$}
\begin{tabular}{llllll} 
\hline
Transit & Planet & Instrument$^c$ & Time (BJD/TDB)$^d$ & Flux$^e$ & Model$^f$ \\
\hline
CoRoT-2b\_20080424\_Mage2v & CoRoT-2b & Mage2v & 2454580.7772671296 & 0.999115 & 1.0 \\
CoRoT-2b\_20080424\_Mage2v & CoRoT-2b & Mage2v & 2454580.7774444446 & 1.002975 & 1.0 \\
CoRoT-2b\_20080424\_Mage2v & CoRoT-2b & Mage2v & 2454580.777630324 & 1.00391 & 1.0 \\
CoRoT-2b\_20080424\_Mage2v & CoRoT-2b & Mage2v & 2454580.777815162 & 1.001988 & 1.0 \\
CoRoT-2b\_20080424\_Mage2v & CoRoT-2b & Mage2v & 2454580.7780003473 & 0.998502 & 1.0 \\
\hline
\multicolumn{6}{l}{\footnotesize{a. CoRoT-2 b, TrES-1 b, WASP-12 b, WASP-19 b, WASP-46 b, and WASP-121 b.}}\\
\multicolumn{6}{l}{\footnotesize{b. A portion is shown for format. The full version is available online.}} \\
\multicolumn{6}{l}{\footnotesize{c.  Ground-based: 23 transits (see \autoref{fig:all_transits}). CoRoT: 82 transits. TESS: 333 transits.}} \\
\multicolumn{6}{l}{\footnotesize{d. Time at mid-exposure. Time system used is  Barycentric Julian Dates (BJD) in Barycentric Dynamical Time (TDB). }} \\
\multicolumn{6}{l}{\footnotesize{e. Normalized detrended flux of planet-hosting star.}} \\
\multicolumn{6}{l}{\footnotesize{f. Model flux for each time in the light curve, using the best fit models for each transit as described in Section~\ref{sec:fit}.}}
\end{tabular}
\label{table:lc_stub}
\end{table}

\subsection{IoIO observations} 
\label{sec:ioio}

The Planetary Science Institute's Io Input/Output Observatory (IoIO) is a small-aperture (35 cm) robotic telescope located at the San Pedro Valley Observatory, a hosting site situated in a dark location 100\,km east of Tucson, Arizona. IoIO was purpose-built to record observations of faint gases around Jupiter that trace their origin from Jupiter's moon Io and as such has a built-in coronagraph \citep[for more details about the observatory, see e.g.][]{morgenthaler19}, but it is also a fully functional general-purpose telescope. Since 2022, much of the time when Jupiter is not available has been dedicated to exoplanet transits, and we have successfully observed over 100 full or partial exoplanet transits, six of whose light curves are published in this work. Observations in 2022-2023 were scripted to be taken in the \textit{R} filter at an exposure time chosen for the magnitude of the host star. For two transits of WASP-12 b, bad weather or scheduling issues resulted in only partial observations over egress.

The time scale for all IoIO observations is determined using the system clock of the observatory control computer, which is maintained to within $\sim$ 3\,ms RMS of UTC using the Network Time Protocol (NTP) system.  However, variable latencies in the Windows 10 operating system, the USB 2.0 link between the computer and Starlight Xpress SX694, and in the camera itself, contribute to a 0.230\,s uncertainty in the observation midpoints for exposures $>$ 0.7\,s (all of the IoIO data presented here) and 0.160\,s for exposures $<=$ 0.7\,s.  This discontinuity has its roots in a decision the MaxIM DL Pro Starlight XPress camera plugin makes about where the exposure time clock is maintained: in the camera (exposures $<=$ 0.7\,s) or within the MaxIm DL software (exposures $>$0.7\,s).  This discontinuity, and thousands of photometric observations of stars recorded since 2019, has made derivation of the quoted uncertainties possible.

Light curves for data taken with IoIO are created through the automatic processing scripts. Photometry is accomplished by creating a segmentation image, using \textsf{photutils} \citep{bradley19}. This approach, more widely used among extragalactic astronomers for isolating galaxies in crowded fields, has the advantage that it is largely agnostic to issues that distort stellar images.  For the IoIO coronagraph, these issues include a variable point-spread-function across the FOV and wind shake.  Standard CCD processing steps, such as bias subtraction, masking of hot pixels and flatfielding, are conducted using a customization of several astropy tools in a multiprocessing, pipeline-oriented code \citep{2022ApJ...935..167A, ccdproc17, morgenthaler22bmp, morgenthaler23ccdmultipipe, morgenthaler23IoIO_code}.  Then, each image is convolved by a 3-pixel FWHM 2D Gaussian kernel to enhance stellar sources.  A preliminary source mask is created to enable optimal background estimation. The critical step in creating the segmentation image is to set the threshold above the local background level that triggers identification of a source.  We use 5 times the RMS variation in the background.  The segmentation image is created assuming a minimum-sized source is formed by 5 connected pixels. Blended sources within the image are separated using a combination of multi-thresholding and watershed segmentation, as per the \textsf{photutils} algorithm, with the number of multi-thresholding levels set to 32 and contrast = 0.001. This procedure results in well-identified sources in most cases, even if the stellar images are not round.  The segmentation image is transformed into a mask that is applied to the original image and used together with the calculated background image to extract source parameters, such as centroid and flux. Astrometry on the source centroids is performed using the \textsf{astrometry.net} software \citep{2010AJ....139.1782L} and the resulting source table is recorded in an Astropy Enhanced Character-Separated Variable file (ECSV). We then performed differential photometry using the star counts for the target star and at least one other nearby companion star, as described in Section~\ref{sec:diff_phot}.

\subsection{SAAO 40-inch observations}

Observations were taken on the 40-inch telescope operated by the South African Astronomical Observatory (SAAO) in Sutherland, South Africa, using the Sutherland High-speed Optical Camera (SHOC) facility instruments \citep{2013PASP..125..976C}. These instruments are ideal for exoplanet observations, as they were built to be optimized for high-speed, accurately timed imaging. They employ Andor iXon 888 frame-transfer EMCCD cameras that are triggered by GPS, mounted with a suite of Johnson-Cousins and Sloan filters, as well as an open-filter setting. Observations were mostly taken in $R$, with other details in \autoref{table:obs_settings}. All data were taken with 2x2 binning (0.334 arcsec per superpixel), the 3 MHz conventional mode, and 5.2x gain setting (8.6 electrons/ADU). Biases and twilight flats were taken each night of the observations, and all data were bias-subtracted and flatfielded. For two transits of CoRoT-2 b and one transit of WASP-12 b, bad weather and/or scheduling issues resulted in only partial observations over ingress and/or egress.

The SAAO 40-inch images have a modest 2.85\arcmin\ x 2.85\arcmin\ field of view that typically contains a handful of well-separated companion stars. Target star counts were calculated using the astropy affiliate package \textsf{photutils} \citep{bradley19} for the target and at least one nearby companion star, using circular apertures in a range of sizes (eight apertures between 8-25 px) to provide stellar locations and counts. None of the six targets in this paper are in crowded enough fields to require PSF photometry. We performed differential photometry using the star counts for the target star and at least one other nearby companion star, as described in Section~\ref{sec:diff_phot}.

\subsection{Differential photometry of IoIO and SAAO 40-inch data}
\label{sec:diff_phot}

We fed the lists of star counts (produced as described above) into a common python-based pipeline to find the aperture and combination of comparison stars that produced the best light curve (nominally, the one with the lowest scatter on the out-of-transit baseline). While differential aperture photometry is simple in concept, in practice it is complicated to construct a robust automated pipeline that can efficiently process ground-based data from multiple instruments and spanning a wide range of target star magnitudes, available comparison stars (from empty fields with 1 very faint comparison star to crowded fields with 100+ stars to choose between), different aperture sizes, instrumentation quirks (e.g., target drift, wind shake, or telescope pointing jumps), and variable weather conditions.

The pipeline tracks stars through pointing jumps or slow drifts, and it can be run iteratively to explore the effects of removing one or more comparison stars and/or image frames. Images are removed if they are missing the target star and/or one or more critical comparison stars, or if the stars are streaked due to telescope motion. Importantly, the light-curve generation notebooks and scripts are quasi-automated, meaning that light curves can be automatically run with default settings appropriate to the selected instrument and planet, but each transit light curve may also be manually adjusted to account for conditions unique to that transit. 

For each target star, the pipeline specifies a default range of star brightnesses relative to the target for acceptable comparison stars. It then calculates the ratio of the target star to each potential companion star and rejects those with scatter above a particular threshold (adjustable for each target star). Comparison stars that were not present on every field (typically due to tracking motion or to variable cloud conditions causing stars to either saturate or become too faint) are automatically removed. Other problematic stars, e.g. strongly variable stars, can be excluded for individual transits as needed. There is also a mechanism to identify bad frames in cases where all comparison stars are lost; after omitting those frames, comparison star selection proceeds as above using the remaining good frames. \autoref{table:obs_settings} shows the number of frames and comparison stars used for each light curve. Once the best set of comparison stars has been identified, a normalized transit light curve is produced for each aperture: the counts from the target star are divided by the sum of the counts of the companion stars, then normalized by dividing by the median flux outside of the predicted transit window. The best aperture is the one that minimizes the scatter on the out-of-transit baseline (see \autoref{table:obs_settings}). Sometimes the resulting transit has a slope in the overall flux, often correlated with the airmass of the target star. We leave detrending any such slopes as a free parameter during fitting (Section~\ref{sec:fit}).



\subsection{\textsc{Minerva}-Australis observations and photometry}

One transit of CoRoT-2 b was observed with the \textsc{Minerva}-Australis telescope array \citep{2019PASP..131k5003A}, located at Mt. Kent Observatory, Australia. \textsc{Minerva}-Australis is an array of four identical 0.7-m telescopes linked via fiber feeds to a single high-resolution spectrograph.  The array has been wholly dedicated to radial-velocity follow-up of TESS planet candidates \citep[e.g.,][]{2019A&A...623A.100N, 2021MNRAS.502.3704A, 2022AJ....163...82W}.  Each telescope is now able to operate in photometric mode using a flip mirror that redirects light to cameras located at the Nasmyth focus.  The observations described here were obtained with a ZWO1600 CMOS camera using 4x4 binning, with the images slightly defocused (PSF FWHM was 5.05 pixels or 3.41"). 
The Minerva telescopes log GPS-based time stamps which are then converted to BJD time stamps using the JDUTC to BDJTDB time converter that is part of the barycorrpy package \citep{2018RNAAS...2....4K}.

An instrumental glitch in the middle of the sequence caused a jump with image rotation, followed by slow derotation back to the original alignment, which combined with deteriorating weather resulted in poorer data quality after midtransit. To control for the effect of the rotation, the final light curve was created using a separate Mathematica photometry notebook to do differential circular-aperture photometry on the target star and two nearby, bright comparisons. Stars were centroided on every frame to determine the centers of the apertures. For each star, three hand-selected boxes well outside of the apertures and without stars were used to determine the background value. The final light curve was generated by subtracting the background counts from each star, dividing the target by the average signal of the comparison stars, and normalizing to the median out-of-transit baseline. The lowest baseline scatter was achieved with an aperture 21 pixels in diameter.

\subsection{IRTF and Las Campanas Observatory Observations and Photometry}

Three transits of CoRoT-2 b were observed in 2008 shortly after the planet's discovery was announced, and they were included in \citet{2010PhDT.......243A} but the light curves have not been published until now. Two transits were taken on the 6.5 m Magellan Clay telescope at Las Campanas Observatory (LCO) on 2008 April 24 and 2008 May 15 during the first observing runs to use the updated MagIC-e2v camera \citep[see e.g.,][]{Osip2004, 2011ApJ...728..125A}. MagIC-e2v had 1024x1024 pixels, each 13 micron square, and on Magellan had a field of view of 38\arcsec\ x 38\arcsec\ and a plate scale of 0.037\arcsec\ per pixel unbinned. The timing for these observations came from network computers that were verified each evening to be synced within a second to UTC standard. Note that the partial transit on 2008 Apr 24 was deliberately scheduled right before dawn after two other transits were observed that night and is not due to weather.

The third transit of CoRoT-2 b on 2008 September 7 was observed using the MIT Optical Rapid Imaging System (MORIS) on NASA's 3-m Infrared Telescope Facility \citep[IRTF; ][]{2011PASP..123..461G}. A dichroic was used to direct light $<0.95 \, \mu {\rm m}$ to MORIS through a Thor long-pass red visitor filter with a lower cutoff at 700 nm. The plate scale for MORIS is 0.114\arcsec\ per pixel, with a 60\arcsec\ x 60\arcsec\ field of view. We used the 1 MHz Conventional mode in 2.4x gain setting, which has a read noise of 6 e- per pixel and gain of 1.5e-/ADU. MORIS observations were triggered by GPS and had better than microsecond timing accuracy.  

Additionally, we observed two transits of TrES-1 b using MORIS on the IRTF on 2009 June 28 and 2009 July 1, using the same settings as for CoRoT-2 b, which have not been previously published. Observational details for all of the IRTF and LCO transits are included in \autoref{table:obs_settings}. 

All of the data from 2008-2009 were reduced using a  Mathematica-based differential aperture photometry pipeline described in \citet{2010PhDT.......243A}. We have elected to keep the original photometry, but we have fit new transit light curve models using the fitting method described in Section~\ref{sec:fit}.

\section{Data from Space Telescopes}
\label{sec:space}

In addition to new ground-based observations, we present new fit results to light curves from at least one TESS sector for all six planets and new fits for the CoRoT mission data for CoRoT-2 b. Although all space-based data used to make these light curves is publicly available and should remain that way in perpetuity, the version of the data available to the public may change in the future (as happened for the CoRoT data). The TESS data also required nontrivial processing (smoothing and detrending) to produce these light curves. For these reasons, we consider it important to publish the version of the transit light curves that we fit in this work, so as to make it as easy as possible to reproduce our results and/or to identify the source of problems that may come to light in the future. All transit light curves may be found in a supplementary electronic file, with a stub table for format at \autoref{table:lc_stub}.

\subsection{TESS photometry}


Where available, we used the published transit midtimes for TESS transits found in \citet{2022ApJS..259...62I}. That work required that transits have at least 75\% of the expected number of data points per transit interval, and we have added midtimes for some partial light curves that were omitted from earlier sectors, as well as sectors that have been released since it was published. We fit a total of 193 TESS light curves to each of six planets as follows: CoRoT-2 b (Sector 54: 12, Total: 12); TrES-1 b (Sector 14: 2, Sector 40: 1, Sector 53: 7, Sector 54: 8, Total: 18); WASP-12 b (Sector 43: 3, Sector 44: 20, Sector 45: 21, Total: 44); WASP-19 b (Sector 09: 2, Sector 36: 5, Sector 62: 32, Sector 63: 33, Total: 72); WASP-46 b (Sector 01: 3, Sector 27: 2, Sector 67: 20, Total: 22); WASP-121 b (Sector 07: 1, Sector 34: 1, Sector 61: 20, Total: 22).

We used the python package \textsf{lightkurve} to download and analyze the 120-s cadence data\footnote{All the TESS data used in this paper can be found in MAST: \dataset[https://doi.org/10.17909/t9-nmc8-f686]{https://doi.org/10.17909/t9-nmc8-f686}.}. To condition the TESS data, we first masked out sections containing bad data. Because TESS light curves span the full transit orbital period, they often exhibit more long-term stellar variability than do ground-based data, which only last a few hours. We found detrending for long-period variability to be necessary before fitting. We masked out the time during planetary transit to interpolate a smoothing function for each sector, using a smoothing window between 21 and 501 frames, depending on the cadence and the target, with the built-in "flatten" method in \textsf{lightkurve}. We then divided the full time series by the smoothed light curve and normalized the light curve. We divided each sector into chunks centered on each transit; all of the transits in this paper had sufficient signal-to-noise to be independently fit to a transit light curve model.

\subsection{CoRoT photometry}

For CoRoT-2 b, we downloaded data from the LEGACY data release (version 4)\footnote{
\url{http://idoc-corot.ias.u-psud.fr/sitools/client-user/COROT_N2_PUBLIC_DATA/project-index.html}}. We used the red flux (keyword ``REDFLUX''), which was less noisy than either the green or the blue fluxes, and the frame times given in BJD/TT (keyword ``DATEBARTT''); for an explanation of why the ms-level difference between TT (Terrestrial Time) and TDB (Barycentric Dynamical Time) is not important for our purposes, see the discussion of timing systems in Section~\ref{sec:timing}. 
For CoRoT-2 b, the first three transits (out of 82) in the 150-day sequence were recorded at the longer cadence (512 s), at which point the transit signature was identified by the CoRoT team, and the rest of the data were taken at 32 s. We chose not to smooth and flatten the CoRoT data in the same way as we did the TESS data but instead use a subset of the data in an eight-hour window centered around the predicted transit midtime. The linear term in our fit function accounted for any remaining trends, which were slight.

\begin{table}
\caption{Stellar and planetary parameters and summary of transits available for 43 planets}
\label{table:system_params}
\small
\begin{tabular}{llr rrrr r | rrr} 
\hline
  & \multicolumn{7}{c|}{Value Used$^a$} & \multicolumn{3}{c}{Observations}\\
\hline
Planet & P (d) & a/$R_\star$ & i & $M_p$ & $M_\star$ & $R_\star$ & Age (Gyr) & $N_{tr}$$^b$ & Epochs$^c$ & Years$^d$\\
\hline
\hline
CoRoT-2b & 1.743 & 6.7 & 87.8 & 3.47 & 0.96 & 0.96 & 0.3 & 164 & 3810 & 18\\
HAT-P-23b & 1.21289 & 4.55 & 85.7 & 2.09 & 1.13 & 1.13 & $4.0(1.0)$ & 125 & 4603 & 15\\
HAT-P-36b & 1.32735 & 4.67 & 85.2 & 1.85 & 1.03 & 1.03 & $6.6^{+2.9}_{-1.8}$ & 129 & 3421 & 12\\
HATS-24b & 1.3485 & 4.67 & 86.6 & 2.26 & 1.07 & 1.07 & $3.7^{+2.0}_{-1.8}$ & 54 & 2301 & 8\\
HATS-35b & 1.821 & 4.79 & 86.9 & 1.22 & 1.32 & 1.32 & $2.13(0.51)$ & 22 & 1573 & 8\\
HATS-70b & 1.88824 & 4.17 & 86.7 & 12.9 & 1.78 & 1.78 & $0.81^{+0.50}_{-0.33}$ & 20 & 1798 & 9\\
HIP 65Ab & 0.98097 & 5.29 & 77.2 & 3.21 & 0.78 & 0.78 & $4.1^{+4.3}_{-2.8}$ & 106 & 1919 & 5\\
KELT-16b & 0.96899 & 3.23 & 84.4 & 2.75 & 1.21 & 1.21 & $3.1(0.3)$ & 93 & 3032 & 8\\
KELT-1b & 1.21749 & 3.69 & 86.8 & 27.23 & 1.32 & 1.32 & -- & 33 & 3288 & 11\\
KOI-13b & 1.76359 & 4.5 & 86.8 & 9.28 & 1.72 & 1.72 & 0.5(0.1) & 8 & 2508 & 12\\
KPS-1b & 1.70633 & 6.37 & 83.2 & 1.09 & 0.89 & 0.89 & -- & 54 & 1521 & 7\\
Qatar-10b & 1.64533 & 4.9 & 85.9 & 0.74 & 1.16 & 1.16 & $3.2(1.9)$ & 217 & 1111 & 5\\
Qatar-1b & 1.42002 & 6.25 & 84.1 & 1.29 & 0.84 & 0.84 & $11.6^{+0.60}_{-4.70}$ & 266 & 3241 & 13\\
Qatar-2b & 1.33712 & 6.45 & 89.0 & 2.49 & 0.74 & 0.74 & $1.4(0.3)$ & 75 & 3354 & 12\\
TOI-2046b & 1.49719 & 4.75 & 83.6 & 2.3 & 1.13 & 1.13 & $0.45^{+0.43}_{-0.021}$ & 102 & 883 & 4\\
TOI-2109b & 0.67247 & 2.27 & 70.7 & 5.02 & 1.45 & 1.45 & $1.77(0.88)$ & 97 & 1723 & 3\\
TOI-564b & 1.65114 & 5.32 & 78.4 & 1.46 & 1.0 & 1.0 & 7.3 & 54 & 943 & 4\\
TrES-1b & 3.03007 & 10.52 & 90.0 & 0.84 & 1.04 & 1.04 & $3.7^{+3.4}_{-2.8}$ & 129 & 2293 & 19\\
TrES-2b & 2.47061 & 7.9 & 83.9 & 1.49 & 1.36 & 1.36 & $5.0^{+2.7}_{-2.1}$ & 228 & 2526 & 17\\
TrES-3b & 1.30619 & 6.0 & 82.0 & 1.91 & 0.93 & 0.93 & $0.90^{+2.80}_{-0.80}$ & 387 & 4545 & 16\\
WASP-103b & 0.92555 & 3.01 & 88.2 & 1.49 & 1.22 & 1.22 & $4.0(1.0)$ & 55 & 3959 & 10\\
WASP-104b & 1.75541 & 6.52 & 83.6 & 1.27 & 1.08 & 1.08 & 3 & 124 & 2098 & 10\\
WASP-114b & 1.54877 & 4.29 & 84.0 & 1.77 & 1.29 & 1.29 & $4.3^{+1.4}_{-1.3}$ & 16 & 2218 & 9\\
WASP-121b & 1.27492 & 3.75 & 87.6 & 1.16 & 1.36 & 1.36 & $1.5(1.0)$ & 93 & 2851 & 10\\
WASP-12b & 1.09142 & 3.04 & 83.4 & 1.47 & 1.43 & 1.43 & 2 & 223 & 5077 & 15\\
WASP-135b & 1.40138 & 5.53 & 82.0 & 1.9 & 0.98 & 0.98 & 0.6 & 54 & 3490 & 13\\
WASP-145Ab & 1.76904 & 8.09 & 83.3 & 0.89 & 0.76 & 0.76 & 6.99 & 21 & 1865 & 9\\
WASP-163b & 1.60969 & 5.62 & 85.4 & 1.87 & 0.97 & 0.97 & -- & 15 & 1352 & 6\\
WASP-164b & 1.77714 & 6.5 & 82.7 & 2.13 & 0.95 & 0.95 & 4.08 & 20 & 1675 & 8\\
WASP-173Ab & 1.38665 & 4.78 & 85.2 & 3.69 & 1.05 & 1.05 & $6.78(2.93)$ & 42 & 2150 & 8\\
WASP-18b & 0.94145 & 3.56 & 84.9 & 10.2 & 1.29 & 1.29 & 0.5 & 108 & 12228 & 32\\
WASP-19b & 0.78884 & 3.46 & 78.8 & 1.15 & 0.96 & 0.96 & $5.5^{+8.5}_{-4.5}$ & 244 & 6714 & 15\\
WASP-32b & 2.71866 & 7.8 & 85.3 & 2.63 & 0.72 & 0.72 & 2.22 & 22 & 1850 & 14\\
WASP-33b & 1.21987 & 3.79 & 87.7 & 2.09 & 1.5 & 1.5 & -- & 102 & 4711 & 16\\
WASP-36b & 1.53737 & 5.85 & 83.2 & 2.36 & 1.08 & 1.08 & 2.5 & 93 & 2919 & 12\\
WASP-3b & 1.84684 & 5.0 & 84.2 & 2.43 & 1.62 & 1.62 & 2.1 & 142 & 3205 & 16\\
WASP-43b & 0.81347 & 4.87 & 82.1 & 1.78 & 0.58 & 0.58 & $7.0(7.0)$ & 224 & 5589 & 12\\
WASP-46b & 1.43037 & 5.85 & 82.8 & 1.91 & 0.83 & 0.83 & $9.6^{+3.7}_{-4.2}$ & 142 & 3364 & 13\\
WASP-50b & 1.95509 & 7.53 & 84.7 & 1.47 & 0.89 & 0.89 & 7 & 51 & 2177 & 12\\
WASP-52b & 1.74978 & 7.38 & 85.3 & 0.46 & 0.87 & 0.87 & -- & 153 & 2565 & 12\\
WASP-5b & 1.62843 & 5.37 & 85.6 & 1.58 & 0.96 & 0.96 & 5.6 & 73 & 3851 & 17\\
WASP-64b & 1.57329 & 5.39 & 86.6 & 1.27 & 1.0 & 1.0 & 7 & 80 & 2993 & 13\\
WASP-77Ab & 1.36003 & 5.41 & 89.4 & 1.67 & 0.9 & 0.9 & $6.2^{+4.0}_{-3.5}$ & 50 & 2898 & 11\\
\hline
\multicolumn{11}{l}{$^a$From the Exoplanet Characterization Catalog \citep{2022ExA....53..547K} and NASA Exoplanet Archive.}\\
\multicolumn{11}{l}{$^b$Number of transit midtimes used in this work.}\\
\multicolumn{11}{l}{$^c$Number of epochs between first and last transit midtime.}\\
\multicolumn{11}{l}{$^d$Years spanned by observed transit midtimes.}\\
\end{tabular}
\end{table}

\section{Transit light-curve fitting}
\label{sec:fit}




Each transit light curve was fit using Markov Chain Monte Carlo (MCMC) methods using the publicly available python package \textsf{PyLightcurve} \citep{2016ApJ...832..202T}, which is based on \textsf{Emcee}, an  affine-invariant ensemble sampler \citep{2013PASP..125..306F} and makes use of several other scientific packages and catalogs\footnote{ \textsf{matplotlib}: \citep{Hunter2007...4160265}; ExoTETHyS: \citep{Morello_2020}; Exoplanet Characterisation Catalogue \citep{2022ExA....53..547K}; \textsf{astropy} \citep{2013A&A...558A..33A}; \textsf{scipy}: \citep{2020NatMe..17..261V}; \textsf{numpy}: \url{https://archive.org/details/NumPyBook}}. For each fit we used 400 walkers and 150,000 links, with the first 30,000 links as burn-in. Convergence of chains was checked by examining the autocorrelation time as well as the traces and the correlations between parameters for signs of nonconvergence. 

Our focus for this work is on the timing, and most of the transit light curve parameters were fixed to the best-fit parameters from more precise light curves, including the eccentricity ($e$), the inclination ($i)$, and the ratio of the orbital semimajor axis to the stellar radius ($a/R_\star$). We used the values reported in \autoref{table:system_params} taken from the Exoplanet Characterization Catalog, which was developed for the ExoClock project  \citep{2022ExA....53..547K}, with missing or (rarely) inaccurate parameters replaced by values from the the NASA Exoplanet Archive where necessary. Note that for most light curves in this work the signal-to-noise ratio is modest enough that the fitted transit midtime values are quite insensitive to the choice of values for, e.g., $a/R_\star$ or the limb darkening coefficients. We nonetheless verified that the fits were reasonable by checking for abnormal scatter in the residuals from the model fits, especially around ingress and egress where issues with the shape are most evident.  Limb-darkening coefficients were calculated for the appropriate bandpass in pylightcuve using ExoTethys \citep{2020JOSS....5.1834M}. For the six planets that are the focus of this work -- CoRoT-2 b, TrES-1 b, WASP-12 b, WASP-19 b, WASP-46 b, and WASP-121 b  -- we have published all parameters, fitted and fixed, for each light curve that we fit, including those from CoRoT and TESS, which are available as a supplementary table with a stub table for format at \autoref{table:lc_fit_params}. We are also publishing the transit midtimes and errors as a standalone table, along with their timing systems and sources, to aid future timing analysis efforts  (\autoref{table:all_midtimes_and_oc}).

We fit each light curve individually with a default of four free parameters -- $T_{mid}$, $r_p/R_\star$, $N$, and $L$ -- where $r_p$ is the planetary radius, $R_\star$ is the stellar radius, $N$ is the normalization constant (typically very close to one), and $L$ is a linear term to account for any residual slopes in the light curve. Some transits with a photometric discontinuity were fit jointly as two partial transits with the same radius ratio and midtime but separate $N$ and $L$ values. For partial transits, we did not fit for $L$. We examined fitting all light curves for a given planet together jointly; however, given the number of fixed parameters, joint fitting would only be helpful for multiple light curves observed with the same filter, which would be expected to share a common value for $r_p/R_\star$. In practice, joint fits were time-consuming to rerun each time new observations were added and did not yield significantly different results, so we have used individual transit fits in this work. 

\begin{table}
\caption{Fit Parameters for Transit Light Curves of Six Planets$^{a,b}$}
\centering
\small
\begin{tabular}{llll} 
\hline
Parameter & Value & Err minus$^c$ & Err plus\\
\hline
Planet & CoRoT-2b &  & \\
Transit number & 0 &  & \\
Transit name & {\rm CoRoT-2b\_20070517-00\_CoRoT} &  & \\
$T_{mid}$ [BJD/TDB] & 2454237.5345 & 0.0017 & 0.0013\\
$a/R_\star$ & 6.7 &  & \\
$i$ (deg) & 87.84 &  & \\
$e$ & 0.0 &  & \\
$\omega$ & 0.0 &  & \\
ldc1$^c$ & 0.5179 &  & \\
ldc2$^c$ & -0.1524 &  & \\
ldc3$^c$ & 0.6843 &  & \\
ldc4$^c$ & -0.3178 &  & \\
$N$ & 0.9988 & 0.00034 & 0.00039\\
$L$ & 0.019 & 0.0019 & 0.0019\\
\hline
\hline
\multicolumn{4}{l}{\footnotesize{a. CoRoT-2 b, TrES-1 b, WASP-12 b, WASP-19 b, WASP-46 b, and WASP-121 b.}}\\
\multicolumn{4}{l}{\footnotesize{b. Stub table for format (see online version for parameters for all light-curve fits).}} \\
\multicolumn{4}{l}{\footnotesize{c. Parameters with no errors were fixed in the fits.}}\\
\multicolumn{4}{l}{\footnotesize{d. Limb-darkening coefficients from ExoTethys 
\citep{2020JOSS....5.1834M}.}} 
\end{tabular}
\label{table:lc_fit_params}
\end{table}

\begin{table}
\caption{Midtime and O-C Values for All Transits Used for Six Planets$^a$}
\begin{tabular}{lllllll} 
\hline
Planet  &	Epoch &	Time system &	$T_{mid}$  &	$T_{mid}$ err &	O-C &	Source\\ 
  &	 &	 &	  &	 [d] &	[d] &	\\ 
\hline
CoRoT-2b & 	-594 & 	BJD\_TDB & 	2453566.49875 & 	0.003 & 	0.01658384 & 	2022ApJS..259...62I; 2010AJ....139...53R\\ 
CoRoT-2b & 	-582 & 	BJD\_TDB & 	2453587.39975 & 	0.003 & 	0.00161869 & 	2022ApJS..259...62I; 2010AJ....139...53R\\ 
CoRoT-2b & 	-209 & 	BJD\_TDB & 	2454237.5345 & 	0.0017 & 	-0.00154782 & 	This work\\ 
CoRoT-2b & 	-208 & 	BJD\_TDB & 	2454239.27859 & 	0.00029 & 	-0.00045492 & 	This work\\ 
CoRoT-2b & 	-207 & 	BJD\_TDB & 	2454241.0222 & 	0.00024 & 	0.00015799 & 	This work\\ 
\hline
\multicolumn{7}{l}{\footnotesize{a. Midtimes for CoRoT-2 b, TrES-1 b, WASP-12 b, WASP-19 b, WASP-46 b, and WASP-121 b. Full table is available online. }} \\

\end{tabular}
\label{table:all_midtimes_and_oc}
\end{table}

\begin{figure}
    \centering
    \includegraphics[width=\textwidth]{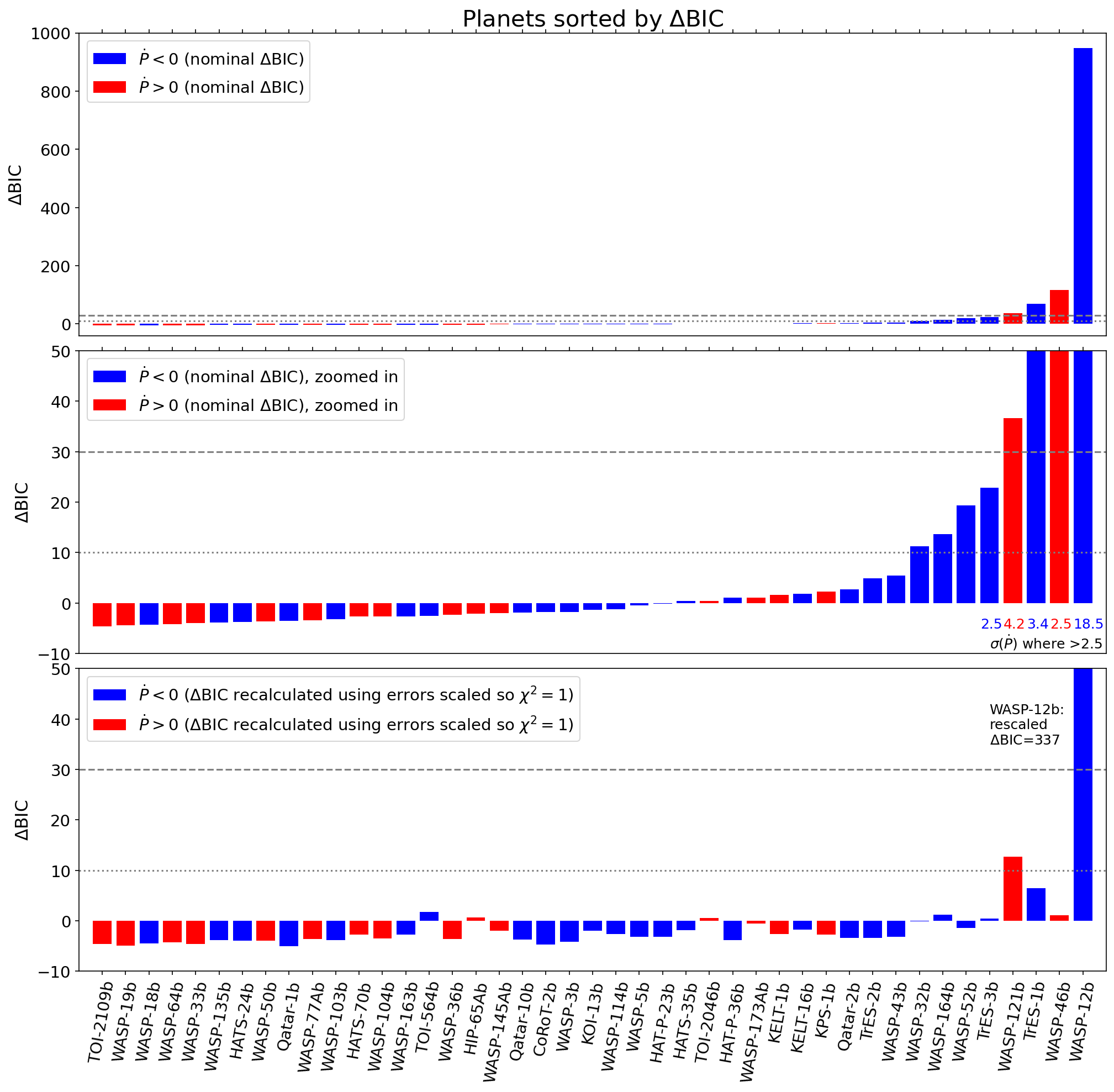}
    \caption{$\Delta {\rm BIC}$ values for 43 planets for which we report at least one new light curve. All panels: Positive $\Delta {\rm BIC}$ values indicate that a quadratic model is preferred over a linear model. Negative values of $\dot P$ (decreasing orbital periods) are shown in blue and positive $\dot P$ (increasing orbital periods) in red. Two horizontal lines show arbitrary cutoffs at $\Delta {\rm BIC}$ = 30 (dashed) and 10 (dotted). Top panel: nominal $\Delta {\rm BIC}$ values (\autoref{table:decay_params}) with WASP-12 b clearly standing apart from the other planets. Middle panel: same data as top, but zoomed in to show all four systems with $\Delta {\rm BIC} \ge 30$ (WASP-121 b, TrES-1 b, WASP-46 b, and WASP-12 b). These four systems also all have at least marginally significant ($\ge 2.5\sigma$) values for $\dot P$, with $\sigma$ values plotted below the bar. Bottom panel: recalculated values of $\Delta {\rm BIC}$ using error bars on all points that have been uniformly rescaled so that the $\chi^2$ value of the fit is equal to 1 (see description of the rescaling test in Section \ref{sec:calc_eph}). Few planets retain positive $\Delta {\rm BIC}$ values, indicating that in many cases high nominal $\Delta {\rm BIC}$ values may be due to underestimated errors.
    }
    \label{fig:sorted_by_delta_bic_all}
\end{figure}

\section{Curation of Literature Light-Curve Times}
\label{sec:lit-data}

Although regular new observations are key to a long-term project such as this one, it is the entire cumulative body of transit knowledge that allows us to make sensitive timing measurements. For several targets, the literature data now span two decades, and some planets have 10+ reference works containing hundreds of transits observed on many telescopes using a variety of methods to generate and fit the transit light curves. Since most of these works contain a mix of new data, reanalyzed data, and times drawn directly or with some correction from the literature, data curation has become and will remain a large component of precise timing analyses. We discuss below some of the lessons learned, including the importance of identifying duplicated times (Section~\ref{sec:dupes}) and composite times (Section~\ref{sec:composites}), and the ever-present pitfall of improper conversion from the JD/UTC and BJD/TDB timing systems (Section~\ref{sec:timing}).

As our starting point for compiling the literature midtimes, we used the work of \citet{2022ApJS..259...62I}, which contains both new analyses of TESS light curves and a thorough literature review through 2022 for hundreds of hot Jupiters. Although we have identified some errors in \citet{2022ApJS..259...62I} that are discussed below, as is perhaps inevitable for any static database of this magnitude, we note that their work represents a formidable contribution to the timing literature. In particular, the table in \citet{2022ApJS..259...62I} has two extremely useful features for error corrections: (1) it converts the literature times into the BJD/TDB time system wherever possible and explicitly states the time system used, and (2) it provides the Astrophysics Data System (ADS) code for the paper where each transit midtime was reported, making it easier to identify the source of every reported midtime. Without these features it would have been much harder to uncover the errors that we have found.

Our second main source of transit midtimes is the Exoplanet Transit Database, or ETD\footnote{\url{http://var2.astro.cz/ETD/}}, which provides an online repository for light curves from amateur astronomers, who have posted tens of thousands of light curves of hundreds of systems since the website went online in September 2008 \citep{2010NewA...15..297P}. Although the data on the ETD can vary widely in quality, many planets have been observed with nearly continuous timing coverage, a key feature for constraining long-term timing variations. For each planet, the ETD provides a table listing all of their available midtimes, which we used with three additional cuts to the data: (1) for systems with many transits, we selected their highest quality data flag (DQ=1 for CoRoT-2 b, TrES-1b,  WASP-12 b, and WASP-19 b, and DQ=1 or 2 for WASP-46 b and WASP-121 b); (2) we removed all transits with very large midtime errors ($>5$ minutes); and (3) we removed duplicate times that are in the ETD but are also in \citet{2022ApJS..259...62I}, preferring the values from \citet{2022ApJS..259...62I}, which are in the BJD/TDB system (the ETD uses HJD/UT; see discussion of timing systems in Section~\ref{sec:timing}). Looking for duplicate transits had to be done by hand to ensure that we did not inadvertently remove points that were observed on the same night but at two different sites and will now be discussed in more detail.

\subsection{Multiple reported transit midtimes for the same epoch}
\label{sec:dupes}

When results are combined from many sources, it is common for there to be multiple, nonidentical transit midtimes reported at the same transit epoch. The challenge is to identify which are duplicates that need to be weeded out and which are simultaneous observations that should be kept. We have found at least four ways in which repeats occur: (1) two or more telescopes were used to observe during the same transit epoch; (2) a single telescope simultaneously observed a transit in two or more wavelengths, producing multiple transit light curves; (3) two or more groups used the same observations to generate different transit light curves and/or made different model fits; and (4) a transit midtime was reported in one timing system (e.g., HJD/UTC) in one work and then was republished later in another timing system (e.g., BJD/TDB). In the first two cases, all of the reported midtimes are independent observations and should be included in timing analyses, but in cases 3 and 4 only one midtime value should be used. Special care needs to be taken with systems that have been repeatedly analyzed, since compilations of compilations often obscure both the source and the original value of the reported transit midtimes.

We checked the midtimes against their references in \citet{2022ApJS..259...62I} for 40 of the 43 UHJs in our full sample (HIP 65 A b, KELT-1 b, and KOI-13 b were not in that work), and found two places where there were duplicated midtimes:

\begin{enumerate}
    \item For WASP-43 b, each transit from \citet{2021ApJS..255...15W} that also appeared in \citet{2020AJ....159..150P} and either \citet{2012A&A...542A...4G} or \citet{2014A&A...563A..41M} shows up twice in \citet{2022ApJS..259...62I}. Since the duplicated transits predate any of the new observations listed in  Table 1 in \citet{2021ApJS..255...15W}, the listed times must have been sourced from one of the two prior works. Removing the 24 duplicate times from \citet{2021ApJS..255...15W} caused the $\Delta {\rm BIC}$ value to fall by half, and the best-fit curvature went from a tiny but 3$\sigma$ positive trend with $\dot P = \left(3\pm1\right)$  ms yr$^{-1}$ to less than 1$\sigma$ from zero (see \autoref{table:decay_params}).
    
    \item For WASP-104 b every transit in both \citet{2021MNRAS.500.5420C} and \citet{2021ApJS..255...15W} was repeated in the table in \citet{2022ApJS..259...62I}. Six transits were removed, and in this case $\Delta {\rm BIC}$ did not change appreciably.
\end{enumerate}

Given the large number of works involved, it is likely there are other instances of duplication in \citet{2022ApJS..259...62I}, especially when citing works such as \citet{2021ApJS..255...15W} which do not list the sources of individual transit midtimes. It is important to note that while the impact on $\Delta {\rm BIC}$ of including duplicates is often small, it can be substantial if the duplicated point has very low errors or occurs at a critical point in the sequence. Duplications are also an entirely unnecessary source of error that could be eliminated with better data management practices by the field as a whole. It is recommended that researchers carefully investigate the source of each transit time used in future analyses, particularly before making claims of subtle detections of small effects. For a long-term approach to how the transit timing community could deal with this problem, see our recommendations in Section~\ref{sec:database}.

\subsection{Composite transit times}
\label{sec:composites}

We now turn to the related issue of composite transit times, or times that are derived from more than one transit light curve. We have identified at least three potential issues that may arise when attempting to use times derived from more than one transit. (1) Composite times typically have error bars that are much smaller than any individual transit, and they compress all of the timing information into a single point, which can distort $\Delta {\rm BIC}$ analyses of long-term variations. (2) Composite times obscure the source of the original timing and make error detection and correction difficult. (3) Composite times may lead to double-counting transits, if one or more of the individual transits that were used to calculate the composite time were also published on their own.

There are three different ways to produce a composite light curve, some of which may be used in timing analyses with suitable precautions and others that should be avoided. We refer to the three main types as a \emph{stacked midtime}, an \emph{ephemeris midtime} (commonly abbreviated as $T_0$), and a \emph{weighted-average midtime}.

\begin{enumerate}
    \item \emph{Stacked transits} result from combining photometric time-series data spanning multiple transit epochs into a single transit light curve. Long photometric time series often have a low signal-to-noise ratio or low sampling for individual transits, and it may not be possible to fit, or even detect, individual transits. By folding on the best-fit orbital period and stacking the data, it is possible to extract a single, higher-quality light curve. Stacked transits are common in discovery papers from large ground-based surveys, as well as for detecting very small planets in Kepler/K2 data \citep[e.g.,][]{2021PSJ.....2..152A}. Kepler-1658 b, for instance, is only robustly detected by stacking all transits within an individual Kepler quarter or TESS sector \citep{Vissapragada_2022}. In principle, there is no issue with using a midtime that was fit to a stacked transit, provided the data being stacked came from a single origin (e.g, the same ground-based survey) and have roughly consistent noise levels and systematics. It is common practice to assign the midtime of the stacked transit to either the first transit epoch in the sequence or the transit epoch nearest to the sequence midpoint. Often, though, particularly for survey discovery data, the assigned midtime for a stacked transit may not be published, except as it may contribute to the fitted value for an ephemeris midtime ($T_0$), discussed next.

    \item \emph{Ephemeris midtimes}, or $T_0$, are the result of fitting a set of individual transit midtimes to the linear ephemeris in \autoref{eqn:linear} and may be variously called the reference midtime, the time of conjunction, the time of transit, or similar. In some cases, $T_0$ is the only timing information provided in a published work. Although it should be common practice to publish the photometric time series for each transit light curve, many papers still do not do so. Nor do papers always report fitted midtimes for each transit light curve. For systems with sparse data, an ephemeris midtime may be one of the few timing points available, and frequently the midtime at the very first epoch is actually an ephemeris midtime. However, using a $T_0$ value in combination with midtimes derived from individual transits can be problematic. A $T_0$ value based on multiple transits taken over many months or years will have a much smaller error bar than any of its component transits, or indeed any individual transit, and thus will serve to anchor the model to the assigned transit epoch. In many cases, this results in a larger value for  $\Delta {\rm BIC}$ than would result if the individual transit midtimes and errors had been used instead. 

    \item The third type of composite time is a \emph{weighted-average} time, where two or more midtimes from independent observations (say, transits at two different telescopes) are averaged together. We do not recommend publishing or using weighted-average times, since timing models can easily deal with multiple data points at the same epoch. Moreover, it is impossible to reconstruct an individual midtime and error from a weighted average, which also makes it impossible to assess for mistakes in timing system conversion or sometimes even to determine which transits were used to calculate the average time. 
\end{enumerate}


\subsubsection{Some examples of composite times in this work}

In Section~\ref{sec:wasp121} we discuss WASP-121 b, where a single composite point, in this case an ephemeris time, is primarily responsible for the nonzero $\Delta {\rm BIC}$ value. In that case, we do not have access to either the individual times or the associated light curves that were used to calculate $T_0$, making its impact hard to evaluate. 

CoRoT-2 b, meanwhile, contains an example of a stacked midtime. The time series from the CoRoT satellite mission encompasses 82 consecutive, high-quality individual transits, but the data were first published as a single stacked midtime assigned to the midtime of the first transit, with $T_{mid} = 2454237.53562\pm0.00014$ [BJD] \citep{2008A&A...482L..21A}. In this work we reanalyzed the original photometry so that we could individually fit each of the 82 transits (see \autoref{table:all_midtimes_and_oc}), which had an average error of -0.00020, +0.00023 days in the timing of each transit (50\% larger than the error assigned to the single stacked midtime). In an experiment, we found significant shifts in the preferred ephemeris model using the stacked midtime rather than the 82 individual midtimes: $\Delta {\rm BIC}$ fell from 33 to $-2$ and $\dot P$ went from 2-$\sigma$ positive curvature to negative curvature that is indistinguishable from 0. In this particular case, since each of the individual transits could be well fit to a light curve model on its own, there is no need to use a stacked transit midtime.

We also found that the CoRoT-2 b mission data were being double-counted in \citet{2022ApJS..259...62I}. Both the stacked composite time (derived from all 82 transits) from \citet{2008A&A...482L..21A} and 79 of the 82 individual transit light curves are included in the table in \citet{2022ApJS..259...62I}. This duplication had two causes: (1) The individual transit midtimes came from \citet{2019MNRAS.486.2290O}, who did not analyze the first three transits taken at the longer 512 s cadence; but since the stacked transit was assigned to the midtime of the first (omitted) transit, it appeared to be at a unique epoch. (2) A typo in \citet{2022ApJS..259...62I} caused the omission of the ``$>1$'' tag to indicate that the \citet{2008A&A...482L..21A} time was derived from more than one transit. In this case, double-counting the CoRoT data had only a modest downward effect on $\Delta {\rm BIC}$ (decreasing it by 6).

\subsection{Some words of caution on timing systems}
\label{sec:timing}

\citet{2010PASP..122..935E} provide a thorough discussion of the various timing systems in use in modern astronomy. Briefly, to remove the effects of both the motion of the Earth around the Sun and the motion of the Sun around the solar system barycenter, a dynamical time system must be used to avoid introducing cyclical variations in an astronomical time series. The recommended standard for reporting exoplanet transit times is Barycentric Julian Dates (BJD) using the Barycentric Dynamical Time (TDB) scale. Critically, for precise timing analyses, all times must be converted into the same timing system before analysis so as to avoid introducing systematic errors. 

Most observational times are originally recorded in JD format in the UTC time scale, hereafter referred to as ``JD/UTC'' and must be converted to barycentric dates, hereafter ``BJD/TDB,'' using the coordinates of the observatory and the coordinates of the star. However, other timing systems are still in common usage. In particular, the heliocenter has historically been easier to calculate, and many midtimes and transit light curves have been reported in Heliocentric Julian Dates (HJD) in the UTC time scale (``HJD/UTC''). Notably, the ETD still uses HJD/UTC for their summary tables. It is also possible to run across times in the Terrestrial Time (TT) system, usually HJD/TT. Both TT and TDB are dynamical time scales with subtly different definitions, and midtimes in BJD/TT differ from midtimes in BJD/TDB by only fractions of a second, which is much smaller than the measured midtime error of any of our transits (or any transit published anywhere in the literature to date). TT and TDB are treated as functionally equivalent for this work.

The difference between the solar system barycenter and the heliocenter is small \citep[no more than a few seconds;][]{2010PASP..122..935E} and BJD and HJD have sometimes been used interchangeably, since almost no transit midtimes are known to better than a few seconds. This practice would reasonable if the times were also both in the same dynamical system -- but BJD times are almost always given in the TDB system, while HJD times are usually (though not always) in the UTC system. The difference between UTC and TDB is currently 69.184 s (in 2024). The UTC-TDB offset changes with the irregular addition of leap seconds that depends on the precise rotation of the Earth, and the last leap second was added on 2016 December 31\footnote{\url{https://www.nist.gov/pml/time-and-frequency-division/time-realization/leap-seconds}}. Thus it is critical to report the full time system used, and herein lie some difficulties with literature light curves. Often, times are reported as BJD only, with BJD/TDB implied, though occasionally light curves will be reported in BJD/UTC. The table of transits from \citet{2022ApJS..259...62I} reports the timing system used for each light curve, usually either BJD/TDB or BJD without any scale specified. However, not all of the works cited by \citet{2022ApJS..259...62I} have properly identified their timing systems, nor have they always converted between timing systems accurately. We have identified multiple errors in this work (see Sections~\ref{sec:corot2b} and ~\ref{sec:wasp19}) relating to missing or erroneously applied BJD/TDB corrections for data that were compiled by \citet{2022ApJS..259...62I}.

For all of our new light curves, we begin with data from the instrument in JD/UTC times and use the \textsf{astropy.time} package to convert to BJD/TDB directly. For ETD light curves, we assume that the times listed in the summary table for each planet have been properly converted to HJD/UTC and apply the correction from UTC to TDB using the same astropy functions. It would be preferable to use the JD times directly from the ETD and convert to BJD, but it is currently not possible to download JD times and observatory coordinates in bulk, and it is impractical to do so by hand for thousands of light curve times. Thus while most of our times are in BJD/TDB, the ETD times are in a hybrid HJD/TDB system. This has minimal impact given the size of the ETD midtime errors.

\subsection{Calculating the best linear and quadratic ephemerides for each planet}
\label{sec:calc_eph}

\begin{figure}
    \centering
    \includegraphics[width=\textwidth]{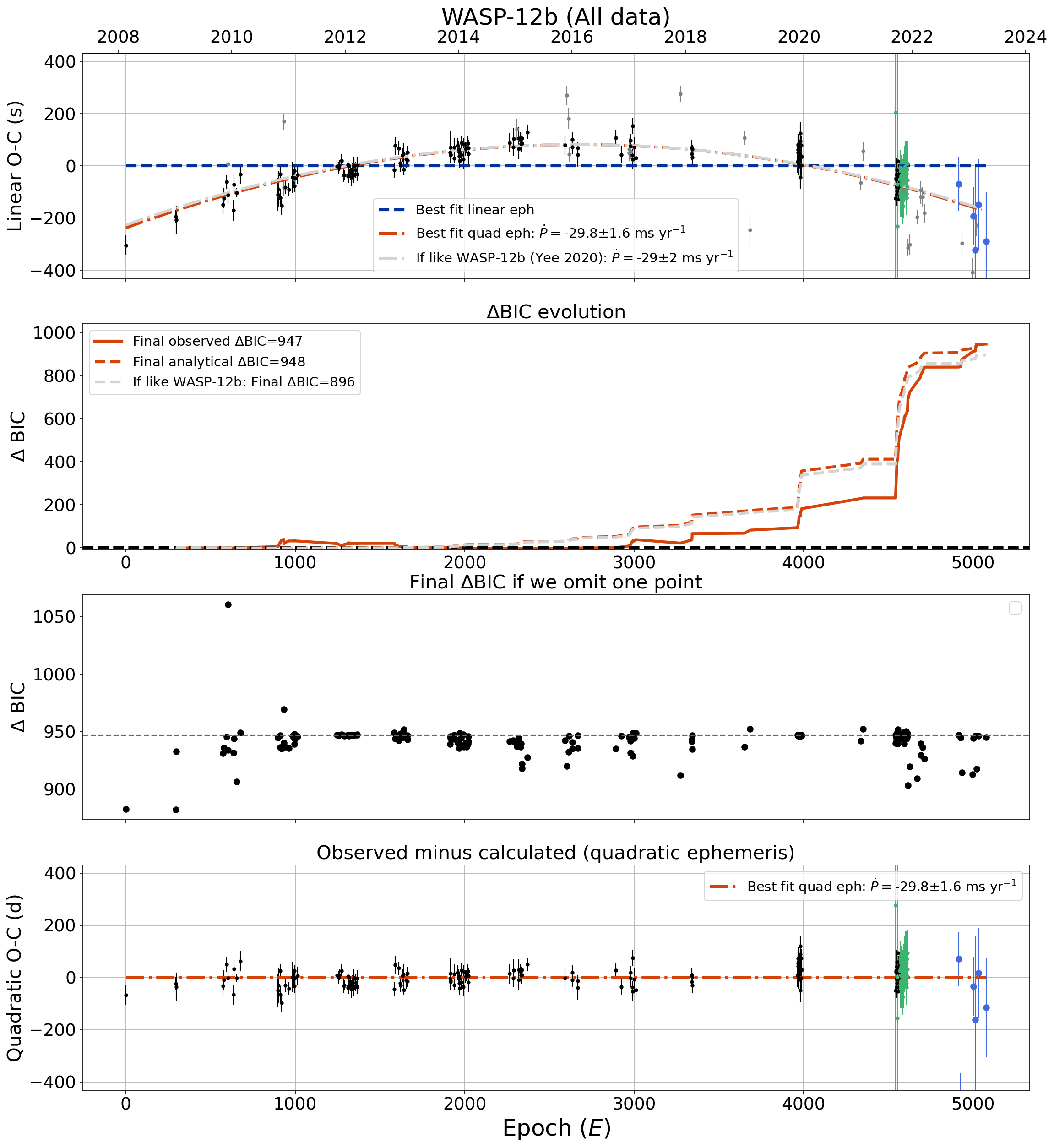}
    \caption{Timing results for WASP-12 b. Panels are listed from top. (Panel 1): observed minus calculated times assuming a linear ephemeris. Midtimes derived from light curves that we fitted are in green (TESS) and blue (new ground-based observations). Literature midtimes compiled by \citet{2022ApJS..259...62I} are in black, while those from the ETD are in grey. If there were critical points from the third panel they would also appear as red diamonds. (Panel 2): $\Delta {\rm BIC}$ from observations (solid orange line) and from analytical calculation assuming our value of $\dot P$ (dashed orange line) \citep[see][]{2023AJ....166..142J}. The expected evolution of $\Delta {\rm BIC}$ if $\dot P = -29.7\,{\rm ms\ yr^{-1}}$, as for WASP-12 b in \citet{2020ApJ...888L...5Y}, is shown in the grey dashed line, with its final value noted in the legend. (Panel 3): final $\Delta {\rm BIC}$ using all data (dashed orange line), compared to the final $\Delta {\rm BIC}$ if just one point were omitted. If there are any points that lead to large shifts (final $\Delta {\rm BIC}$ changes by $\ge25\%$ and at least 5) they are highlighted with red diamonds in panels one and three. (Panel 4): observed minus calculated times for the best-fit quadratic ephemeris. }
    \label{fig:wasp12b_timing}
\end{figure}

Because of the complexity of the datasets involved, our timing analysis took an iterative approach as we identified errors in the source data. We began with the series of transit midtimes and errors from both our own fits (\autoref{sec:fit}) and our curated collection of literature times (\autoref{sec:lit-data}). For each system, we then ran the following sequence of calculations:

\begin{enumerate}
    \item Fit the best linear ephemeris (\autoref{eqn:linear}) and quadratic ephemeris (\autoref{eqn:quad}) to the available midtimes. 
    
    \item Calculated the $\Delta {\rm BIC}$ value (\autoref{eqn:deltaBIC}) to see which model is preferred. Negative values of $\Delta {\rm BIC}$ indicate a linear ephemeris is preferred, while positive values indicate a quadratic ephemeris is preferred. The sign of $dP/dE$ determines if the system had a period increase (positive $dP/dE$) or decrease (negative $dP/dE$). Values of $\Delta {\rm BIC}$ for each planet are shown in \autoref{fig:sorted_by_delta_bic_all}.
    
    \item Ran a series of omit-one tests, omitting each individual midtime in turn and calculating the resulting $\Delta {\rm BIC}$. Erroneous mid-transit times can significantly impact $\Delta {\rm BIC}$, particularly if they have very small error bars that act to anchor to a particular model. Individual points whose absence or presence changes $\Delta {\rm BIC}$ by more than 25\% were flagged and show up as red diamonds in Figures \ref{fig:wasp12b_timing} to \ref{fig:kepler-1658b}. 

    \item Calculated the rescaled $\Delta {\rm BIC}$ value. Some, though not all, transit midtimes are reported with unrealistically low error bars. The ideal practice would be to refit all literature light curves with the same fit model and method; however, this is usually impractical or impossible, especially for large datasets where the photometry for many light curves has not been published. Instead, we devised a "rescaling test", where we used the $\chi^2$ value of the best linear fit using the original errors and then scaled all error bars up by the same factor, so that $\chi^2=1$, then ran a separate fit to the rescaled data. We then calculated the resulting $\Delta {\rm BIC}$ value using the rescaled error bars. \autoref{fig:sorted_by_delta_bic_all} shows the impact of rescaling, which almost always decreases the value for $\Delta {\rm BIC}$, even causing some values to become negative (switching to a preference for a linear ephemeris); however, genuine detections such as WASP-12 b remain highly significant (rescaled $\Delta {\rm BIC}=331$ versus original $\Delta {\rm BIC}=936$). 
    
    \item Investigated outliers, especially those that failed the omit-one test. If any times needed to be removed or altered, we then reran the above.
\end{enumerate}

\subsection{Identifying Systems of Interest}

We show the $\Delta {\rm BIC}$ values for all 43 systems for which we have observed at least one new transit in \autoref{fig:sorted_by_delta_bic_all}. We chose $\Delta {\rm BIC}>30$ as a somewhat arbitrary threshold to investigate a small number of systems with the highest likelihood of a changing ephemeris. (See Section \ref{sec:marginal} for a discussion of this threshold.) We discuss below the timing results for four planets, including WASP-12 b, that still meet our $\Delta {\rm BIC}>30$ criteria, as well as two planets, CoRoT-2 b and WASP-19 b,  whose initially promising $\Delta {\rm BIC}$ values ultimately proved to result from errors in the literature.

\begin{table}
\caption{Best Fit Transit Ephemerides for Six Planets}
\centering
\small
\begin{tabular}{lllll} 
\hline
\multicolumn{3}{c}{\textbf{Decreasing Orbital Periods}}\\
Parameter & WASP-12b & TrES-1b$^a$\\
\hline
$\Delta {\rm BIC}$ & $947.1$ & $68.8$\\
Linear P (d) & $1.09141892\pm0.00000005$ & $3.03006986\pm0.00000010$\\
Linear $T_{mid}$ & $2457800.69944\pm0.00008$ & $2457868.26528\pm0.00010$\\
Quad. P (d) & $1.09141858\pm0.00000004$ & $3.03006915\pm0.00000023$\\
Quad. $T_{mid}$ & $2457800.70032\pm0.00007$ & $2457868.26561\pm0.00014$\\
Quad. dP/dE & ($-1.031\pm0.056$)$\times10^{-11}$ & ($-1.576\pm0.47)$$\times10^{-9}$\\
$\dot P$ [ms yr$^{-1}$] & $-29.8\pm1.6$ & $-16.4\pm4.8$\\
$\sigma$ ($\dot P$) & $18.5$ & $3.4$\\
$Q_\star'$ & $1.6\times 10^5$ & $1.6\times 10^2$\\
Lifetime (Myr) & $3.2$ & $16.0$\\
\hline
\multicolumn{3}{c}{\textbf{Increasing Orbital Periods}}\\
Parameter & WASP-46b & WASP-121b\\
\hline
$\Delta {\rm BIC}$ & $116.8$ & $37.6$\\
Linear P (d) & $1.43037298\pm0.00000014$ & $1.27492474\pm0.00000007$\\
Linear $T_{mid}$ & $2458343.17327\pm0.00020$ & $2459223.80556\pm0.00005$\\
Quad. P (d) & $1.43037343\pm0.00000023$ & $1.27492508\pm0.00000010$\\
Quad. $T_{mid}$ & $2458343.17281\pm0.00027$ & $2459223.80557\pm0.00005$\\
Quad. dP/dE & ($1.027\pm0.42$)$\times10^{-9}$ & ($5.6\pm1.3$)$\times10^{-10}$\\
$\dot P$ [ms yr$^{-1}$] & $22.7\pm9.1$ & $13.9\pm3.3$\\
$\sigma$ ($\dot P$) & $2.5$ & $4.2$\\
\hline
\multicolumn{3}{c}{\textbf{Constant Orbital Periods}}\\
Parameter & CoRoT-2b & WASP-19b\\
\hline
$\Delta {\rm BIC}$ & $-0.8$ & $-4.4$\\
Linear P (d) & $1.74299709\pm0.00000007$ & $0.78883900\pm0.00000001$\\
Linear $T_{mid}$ & $2454580.90647\pm0.00007$ & $2458860.73532\pm0.00004$\\
Quad. P (d)$^b$ & $1.74299728\pm0.00000029$ & $0.78883902\pm0.00000003$\\
Quad. $T_{mid}$$^b$ & $2454580.90650\pm0.00008$ & $2458860.73530\pm0.00005$\\
Quad. dP/dE$^b$ & ($-1.5 \pm2.2$) $\times10^{-10}$ & ($1.4\pm1.9)$ $\times10^{-11}$\\
$\dot P$ [ms yr$^{-1}$] & $-2.8\pm3.9$ & $0.6\pm0.8$\\
$\sigma$ ($\dot P$) & $0.7$ & $0.7$\\
\hline
\hline
\multicolumn{3}{l}{\footnotesize{a. With $Q_\star' < 200$, orbital decay is an unlikely explanation.}} \\
\multicolumn{3}{l}{\footnotesize{b. Best fit quadratic ephemeris for comparison; use linear ephemeris for prediction.}} 
\end{tabular}
\label{table:nonlinear_systems_params}
\end{table}

\section{Timing Analysis Results for Individual Systems}
\label{sec:results}

\subsection{WASP-12 b: solid detection of decreasing orbital period}
\label{sec:wasp12b}

Timing data for the planet with the clearest case for orbital decay, WASP-12 b \citep{2017AJ....154....4P, 2019MNRAS.490.1294B, 2020ApJ...888L...5Y}, with $P=1.09$ d, are shown in \autoref{fig:wasp12b_timing}. Six new full or partial transits were observed with IoIO for this project in 2022-2023 (\autoref{table:obs_settings}). We also fit two sectors of TESS data (Sector 44 and Sector 45) that were released after the analysis of \citet{2022ApJS..259...62I}, which had Sector 20 and Sector 43.  We used 145 light curves from the literature as compiled by \citet{2022ApJS..259...62I}, plus 28 additional, nonduplicated light curves from the ETD \citep[requiring midtime errors less than five minutes and DQ=1,][]{2010NewA...15..297P}. The best-fit parameters for linear and quadratic ephemerides are shown in \autoref{table:nonlinear_systems_params}. Despite the modest precision of the new data (due mostly to unfavorable weather), our value for $\dot P = -29.8\pm1.6$ ms yr$^{-1}$ is quite similar to previously published values, e.g. $-29\pm2$ ms yr$^{-1}$ from \citet{2020ApJ...888L...5Y}.

\begin{figure}
    \centering
    \includegraphics[width=\textwidth]{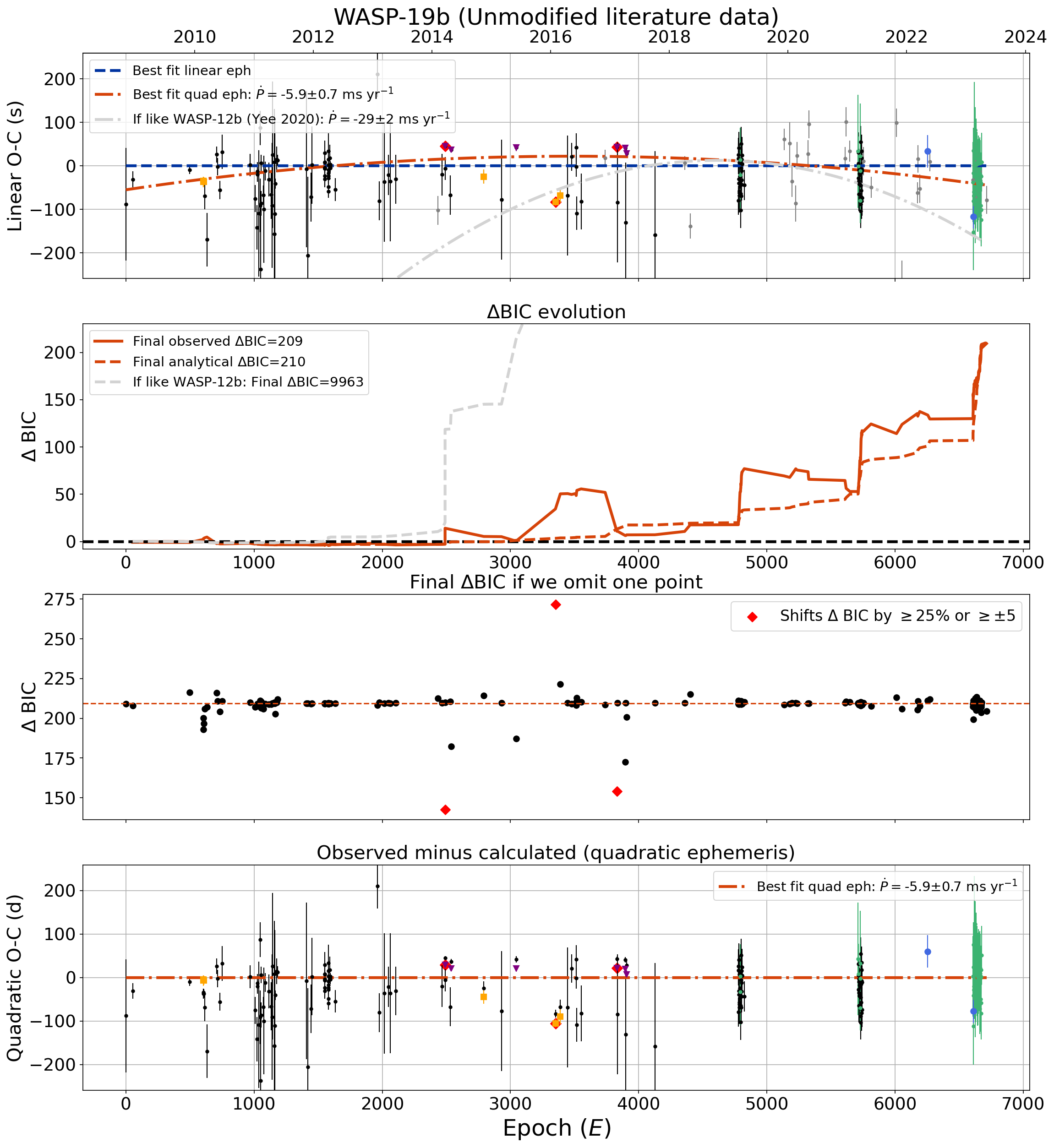}
    \caption{Timing results for WASP-19 b using unmodified literature times, as discussed in Section \ref{sec:wasp19}. The description of the data is the same as \autoref{fig:wasp12b_timing} except that times from three works are highlighted: six transits from \citet{2019MNRAS.482.2065E} which appear as purple triangles, center; three transits from \citet{2013MNRAS.428.3671T} which appear as a single orange square, left; and three transits from \citet{2017Natur.549..238S} which appear as orange squares, center. These works all either have errors or are responsible for critical points that strongly affect $\Delta {\rm BIC}$; see discussion in text.}
    \label{fig:wasp19b_timing_case1}
\end{figure}

\begin{figure}
    \centering
    \includegraphics[width=\textwidth]{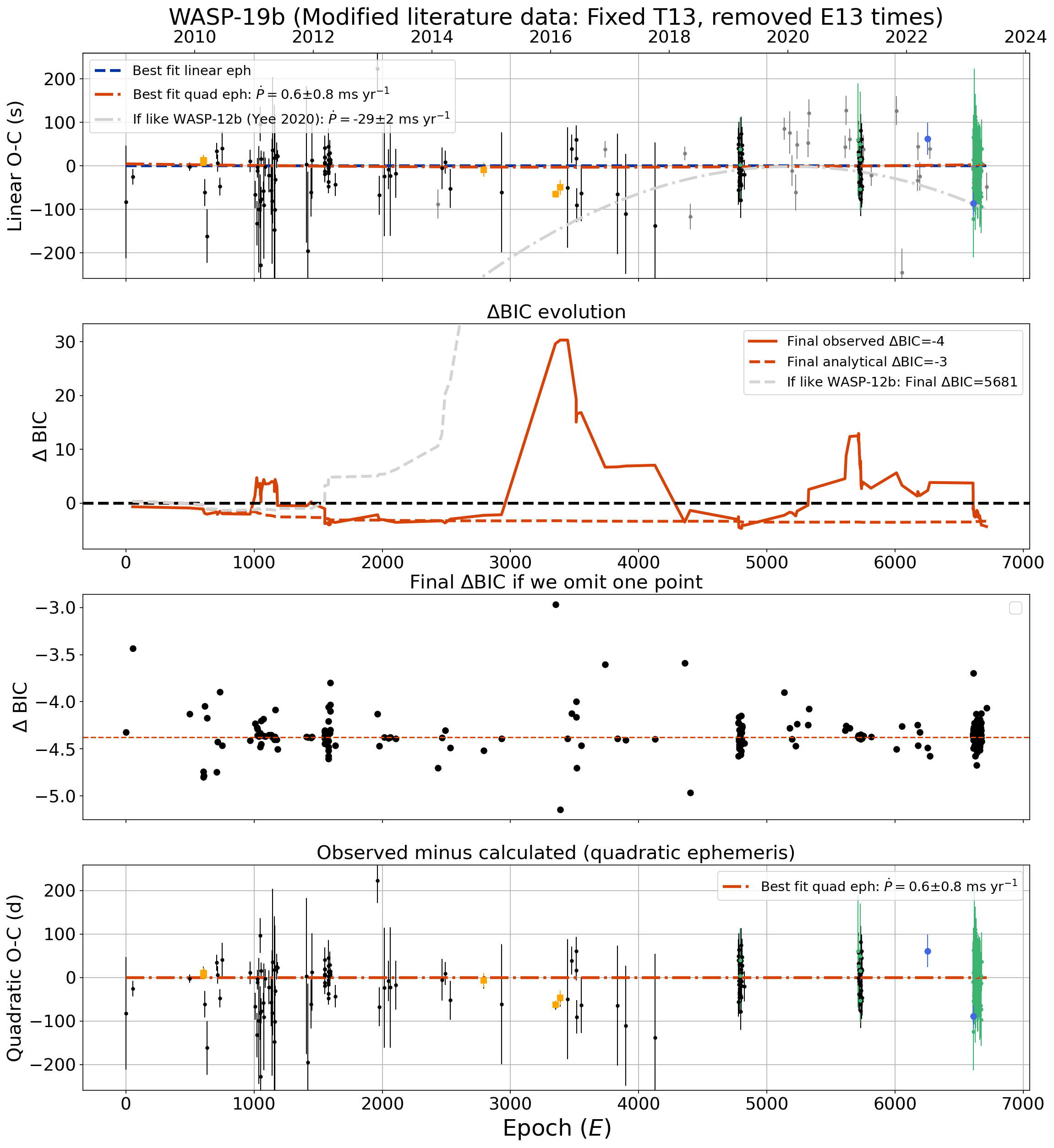}
    \caption{Timing results for WASP-19 b using the modified literature times, as discussed in Section \ref{sec:wasp19}. Description of the data is the same as \autoref{fig:wasp12b_timing}, except that times from two works are highlighted:  three transits from \citet{2013MNRAS.428.3671T} which appear as a single orange square at left, and three transits from \citet{2017Natur.549..238S} which appear as orange squares at center. Note that the timing for three transits from \citet{2013MNRAS.428.3671T} have been corrected and six transits from \citet{2019MNRAS.482.2065E} have been removed, compared to \autoref{fig:wasp19b_timing_case1}. Meanwhile, the times from \citet{2017Natur.549..238S} are no longer critical points that strongly impact the fitted value for  $\Delta {\rm BIC}$. See discussion in the text.}
    \label{fig:wasp19b_timing}
\end{figure}


\subsection{WASP-19 b: no evidence for orbital decay}
\label{sec:wasp19}

WASP-19 b, discovered in \citet{2010ApJ...708..224H}, has one of the shortest orbital periods for a UHJ at just under 0.79 days, and as such has long been viewed as a promising target for orbital decay. It also has one of the longest observational baselines, with data spanning 15 years and nearly 7000 epochs (\autoref{fig:wasp19b_timing_case1}). Only upper limits on orbital decay were found by \citet{2020MNRAS.491.1243P} and \citet{2022A&A...668A.114R}, though a weak $3\sigma$ period decrease detection was reported by \citet{2022ApJS..259...62I}. These prior works have made it clear that if WASP-19 b is decaying, it must be doing so at least an order of magnitude more slowly than WASP-12 b, and consequently $Q_\star'$ must be at least an order of magnitude larger ($\ge10^6$). 

Our analysis includes 2 new transits from the SAAO 40-inch in 2022 as well as two newly-fit sectors of TESS data (Sectors 62 and 63), for a total of 74 new transit midtimes. 

We started with 141 literature times for WASP-19 b listed in \citet{2022ApJS..259...62I}. Two points were rejected for having errors greater than five minutes. We then identified 10 times from \citet{2020A&A...636A..98C} that were not included in \citet{2022ApJS..259...62I}, of which we omitted two: one had midtime error greater then five minutes, and the other is a non-obvious duplicate. The duplicate was found in Table A.5 of \citet{2020A&A...636A..98C}, where two nominally different epochs are reported at E=-2063 and $T_{\rm mid} = 2454775.3372$ from \citet{2010ApJ...708..224H} and  E=-2061 and $T_{\rm mid} = 2454776.91566$ from \citet{2010A&A...513L...3A}. On closer examination the midtime at E=-2061 is derived from a joint analysis of an occultation observed on 2009 May 3 by \citet{2010A&A...513L...3A} and the same transit light curve from 2008 Nov 11 from \citet{2010ApJ...708..224H} that was reported at E=-2063, and thus does not represent an independent midtime measurement. (Occultation timing analysis is beyond the scope of the present work.) We also downloaded 84 midtimes from the ETD (through Feb 2024). Of these, 34 midtimes had DQ=1 and errors less than five minutes \citep{2010NewA...15..297P}. This number was reduced to 29 unique transits after removing two midtimes that were entered from the literature \citep{2010ApJ...708..224H, 2011AJ....142..115D}, two midtimes from lightcurves that were later reanalyzed and published by \citet{2020MNRAS.491.1243P} and then \citet{2022ApJS..259...62I}, and one identical duplicate in the ETD table. Our total number is thus 176 literature midtimes. 

We then made three small modifications to the literature record. (1) We restored an extra significant figure to the values from \citet{2022ApJS..259...62I}, which were rounded to 5 places, using the original values published in \citet{2020MNRAS.491.1243P}. This avoids misidentifying as duplicates a few transits that were observed in multiple filters using the same instrument due to having very similar midtimes but otherwise does not impact results. (2) For one transit originally published by \citet{2019MNRAS.482.2065E}, we made a different choice than \citet{2022ApJS..259...62I} about which of several published midtimes to use. All six transits published by \citet{2019MNRAS.482.2065E} also appear with identical values and errors in \citet{2020AJ....159..150P} after converting to BJD/TDB, but just one appears, non-identically, in \citet{2020MNRAS.491.1243P} where the light curve was reanalyzed; this was the value that appeared in \citet{2022ApJS..259...62I}. The difference between the two midtimes in question is 26 seconds, and \citet{2020MNRAS.491.1243P} also found much higher error bars (13 s vs. 4 s). Since only one of the six \citet{2019MNRAS.482.2065E} transits was reanalyzed by \citet{2020MNRAS.491.1243P}, and since the \citet{2019MNRAS.482.2065E} times are critical points for the $\Delta {\rm BIC}$ analysis (about which more soon) we made the decision to treat the six transits consistently and used all six BJD/TDB midtimes as published in \citet{2020AJ....159..150P}. (3) For the three times from \citet{2017Natur.549..238S}, we swapped in the original values and errors for the ones that \citet{2020MNRAS.491.1243P} published using re-fit light curves and which are used by \citet{2022ApJS..259...62I}. The midtime values were nearly the same, but two of the three light curves have notably smaller error bars in \citet{2020MNRAS.491.1243P}, especially the transit $T_{mid}=2457448.71294 \pm 0.00020$ in the original vs. the refit $T_{mid}=2457448.71292 \pm 0.0000766$ -- less than 2 sec difference in midtime but a factor of three smaller error. We have chosen to use the larger error as the more conservative choice, although the ultimate effect on $\Delta {\rm BIC}$ after correcting the other issues noted below is minimal.

Using the essentially unmodified literature data as just described we find $\Delta {\rm BIC} = 209$, which is quite high, with a statistically significant $\dot P=-5.9\pm0.7$, a nominal 8-sigma detection (see \autoref{fig:wasp19b_timing_case1}). However, closer inspection reveals several issues that bring this result into doubt, which we will now discuss in more detail, along with the impact of each on the calculated $\Delta {\rm BIC}$ value. 

\begin{enumerate}
    \item First, we found an apparent timing conversion error in \citet{2013MNRAS.428.3671T}, which has since been propagated to other works \citep[including][]{2020AJ....159..150P,2022ApJS..259...62I}. Two tables in \citet{2013MNRAS.428.3671T} contain timing information, with their Table 2 listing the fitted midtimes for three transit light curves in HJD/UTC and their Table 4 converting to HJD/TDB for direct comparison with other works. At the observation epoch the difference between these two time systems should be about 66 seconds (accounting for both leap seconds and the TAI constant offset), but the actual differences between the listed midtimes are 41 s, 33 s, and 38 s. The cause for this difference is unknown, but may be an incomplete timing conversion, where either the leap seconds or the TAI offset were included, but not both. Using the online tool from \citet{2010PASP..122..935E}\footnote{\url{https://astroutils.astronomy.osu.edu/time/hjd2bjd.html}. Starting from HJD times carries a disclaimer in that tool:  ``Due to the approximations inherent to most HJD calculators, this conversion cannot be trusted to better than 1 second."} to directly convert the times in their Table 2 from HJD/UTC to BJD/TDB, we find values of 2455251.797060, 2455252.585840, and 2455255.741230, which are used in all subsequent analyses. The uncorrected points are highlighted in \autoref{fig:wasp19b_timing_case1} as the orange squares around E=600.

    Fixing the timing of these three points alone caused a 25\% drop to $\Delta {\rm BIC}=147$.

    \item After correcting the timing for the three times from \citet{2013MNRAS.428.3671T}, the strength of the $\Delta {\rm BIC}$ value rests on three points which fail our omit-one test; removing each point individually changes the $\Delta {\rm BIC}$ value for the system by 30\%. Two of these times are among the six transits discussed above from \citet{2019MNRAS.482.2065E}. These transits were white light transits created by combining high-resolution spectroscopic observations with the 6.5 m Magellan telescope using IMACS, and were of such high precision that even with visible spot crossings during transit the timing precision was between 4-9 s in the original work. Besides the low errors, all six points from \citet{2019MNRAS.482.2065E} appear above the line for the best linear ephemeris by about a minute (see the purple triangles in \autoref{fig:wasp19b_timing_case1}). It is worth noting that while at first glance this might appear to be a UTC/TDB conversion error, we have not found any definitive evidence to support that hypothesis. The values in \citet{2019MNRAS.482.2065E} were published in BJD/UTC, as confirmed by correspondence with the authors of that work, and apparent differences between the values in  \citet{2019MNRAS.482.2065E} and subsequent works are due to correctly converting the original midtimes to BJD/TDB. The source of the apparent universal offset is thus still unknown.

    In combination, the very low errors and the consistent offset are entirely responsible for all of the remaining $\Delta {\rm BIC}$. We explored two approaches to dealing with this issue:

    \begin{enumerate}
     
        \item Increasing the errors on the six points from \citet{2019MNRAS.482.2065E}. Noting that the single transit that was re-fit by \citet{2020MNRAS.491.1243P} had three times the fitted error as that reported by \citet{2019MNRAS.482.2065E}, we tripled the error bars on all six transits and recalculated the value of $\Delta {\rm BIC}$. This caused a drop to insignificance, with $\Delta {\rm BIC}=-3$ and $\dot P=-0.8\pm0.7$. However, one of the \citet{2017Natur.549..238S} points remains an outlier in this approach.

        \item  Removing the six points from \citet{2019MNRAS.482.2065E}. This causes the effect to vanish with $\Delta {\rm BIC}=-4$ and $\dot P=0.6\pm0.8$, a strong preference for a linear ephemeris. Also, with this approach there are no critical times that affect the $\Delta {\rm BIC}$ value by more than 25\%. This is the version we adopted as preferred and which will be used in subsequent analyses in this paper.  With these modifications to the  literature times, we find that decay, if it is happening for WASP-19 b, is proceeding at an as-yet undetectable pace.

    \end{enumerate}
\end{enumerate}

\subsection{CoRoT-2 b: no orbital decay}
\label{sec:corot2b}

\begin{figure}
    \centering
    \includegraphics[width=\textwidth]{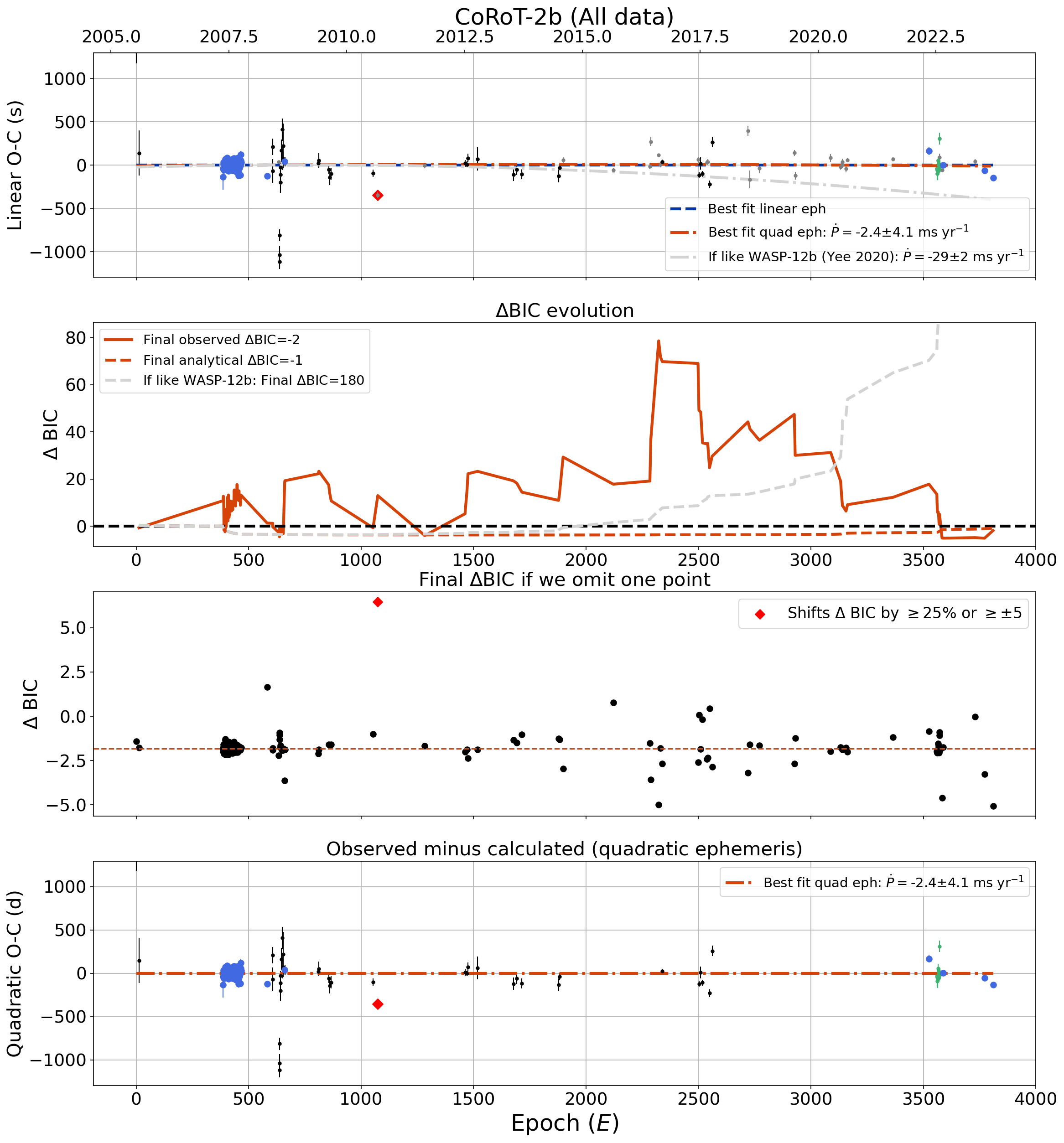}
    \caption{Timing results for CoRoT-2 b with new fit values for CoRoT mission data replacing the values used by \citet{2019MNRAS.486.2290O, 2022ApJS..259...62I}. Instead of a strong negative curvature CoRoT-2 b shows no signs of nonlinearity. Description of data same as \autoref{fig:wasp12b_timing}.}
    \label{fig:corot2b}
\end{figure}

CoRoT-2 b, with $P=1.74$ d, was the second planet announced by the CoRoT space mission \citep{2008A&A...482L..21A}, and it has one of the longest baselines in our survey, with data spanning 18 years (\autoref{table:system_params}). Prediscovery data using BEST show clear indications of transits two years before the CoRoT discovery, though only one of those light curves spanned both ingress and egress \citep{2010AJ....139...53R}. This planet has been frequently studied, since in addition to being one of the earliest known transiting planets, it orbits a highly active \citep[7-20\% of the surface covered in spots,][]{2011A&A...527A..20G}, rapidly rotating ($P_{rot}=4.5$ d), likely young (30-300 Myr) star \citep{2010A&A...511A...3G}, and the planet itself has an inflated radius ($R_p=1.5 R_J$ and $M_p = 3.3 M_J$) that is a challenge to reproduce theoretically \citep{2010A&A...511A...3G}. Previous work has found suggestions of nonlinearity in the ephemeris: \citet{2019MNRAS.486.2290O} measured $\dot P = -8 \pm 3$ ms y$^{-1}$, though with BIC values that favor a linear model, while \citet{2022ApJS..259...62I}, using additional data from TESS, found a much higher $\dot P = -104 \pm  6$ ms y$^{-1}$. In both cases high scatter was seen in the timing residuals. Recently, \citet{2024ApJS..270...14W} found that the tiny error bars reported in \citet{2019MNRAS.486.2290O} strongly influenced the values for $\dot P$; using more realistic errors they found a significantly decreasing period with $\dot P = -22 \pm 3$ ms y$^{-1}$, between the two previous estimates.  

We report seven new full or partial transits: three observed in 2008 \citep{2010PhDT.......243A}, three observed in 2022-2023 with the SAAO 40 inch telescope, and one observed in 2023 with \textsc{Minerva}-Australis. We have also fit 12 TESS transits from Sector 54, which was released after the publication of \citet{2022ApJS..259...62I}.  Finally, we have refit the original CoRoT mission data, which observed 82 consecutive transits between 2007 May and October. We include in our timing analysis the reported transit midtimes for 36 light curves from the literature, as compiled by \citet{2022ApJS..259...62I} after omitting duplicates and one composite time, plus 29 additional, nonduplicated light curves from the ETD, requiring midtime errors less than five minutes and DQ=1 \citep{2010NewA...15..297P}. Such a large and diverse dataset is very useful for highlighting the challenges that arise with long-term timing analyses. We describe the case of CoRoT-2 b in detail to illustrate how more than one kind of error can coexist in the same data set and to illustrate the detective work that is required to both identify and fix the errors.


Our initial examination of the system added the seven new ground-based transits plus TESS Sector 54 to the midtimes collected by \citet{2022ApJS..259...62I}, but did not reanalyze the CoRoT mission times. Using these data, CoRoT-2 b had the second strongest signature for a non-linear orbital period in our sample (after WASP-12 b), with a 3.1-$\sigma$ detection of a decreasing period and a $\Delta {\rm BIC}$ value of 383. The initial best-fit rate of decay was calculated to be $\dot P = -29.4 \pm 9.5$ ms y$^{-1}$. This value was consistent with what \citet{2019MNRAS.486.2290O} reported and would have implied a very low value for $Q_{\star}' < 10^4$, and thus we subjected the data to additional scrutiny.

Three issues were noted right away: 

\begin{enumerate}
    \item \emph{An outlier.} One point from the ETD database caused the $\Delta {\rm BIC}$ value to change by $\pm100$. This was found to be an error in the summary table on the ETD website, which incorrectly reported the JD, rather than HJD, time of observation, as became apparent when consulting the individual transit details page.\footnote{\url{http://var2.astro.cz/EN/tresca/transit-detail.php?id=1343557668}, last accessed 2023 Oct 13.} We have corrected the midtime in our analysis, as well as in \autoref{table:all_midtimes_and_oc}; however, we caution that it is likely that such a large database as the ETD contains other similar errors. 

    \item \emph{Underestimated errors.} The second issue involves the published times for the CoRoT mission data from \citet{2019MNRAS.486.2290O}. The CoRoT mission discovery dataset contains 82 consecutive, high-quality transits at the very beginning of the time series, and are highly influential on the observed $\Delta {\rm BIC}$ value. However, the errors on individual transit midtimes reported in \citet{2022ApJS..259...62I} and originally published in \citet{2019MNRAS.486.2290O} were much smaller than the scatter between midtime residuals (O-C values), indicating that the errors were probably underestimated, as noted by \citet{2024ApJS..270...14}. Indeed, Figure 1 in \citet{2019MNRAS.486.2290O} shows adjusted transit midtime errors that are $\sim10$ times higher, though their table (perhaps inadvertently) published the smaller errors. Low error bars on 79 precise points at an early epoch have a huge impact on the $\Delta {\rm BIC}$ value, so correcting these errors was of critical importance.

    \item \emph{Strong unexplained trend.} The most serious issue was a trend that was apparent by eye in the 150 days of CoRoT mission data from \citet{2019MNRAS.486.2290O}. This trend would have implied a faster rate of decay than that of WASP-12 b and $Q_\star^\prime < 10^4$. In contrast to WASP-12 b, for which the star may be slightly evolved (see Section~\ref{sec:disc}), if anything the evidence points toward CoRoT-2 being a young star, with corresponding high values expected for $Q_\star^\prime$. Removing the CoRoT points entirely caused the $\Delta {\rm BIC}$ value to drop near 0, so it was clear that any apparent detection of orbital period decrease rested on the timing of these transits.

\end{enumerate}

To get to the bottom of both (2) and (3), we downloaded the CoRoT mission data and refit all 82 transits (see Section \ref{sec:lit-data}). Using our new times, which are shown in \autoref{table:all_midtimes_and_oc}, we re-ran the timing analysis and found that $\Delta {\rm BIC}=-2$, indicating no orbital decay. Our timing results are show in \autoref{fig:corot2b}. The differences between our values for the CoRoT mission times and the midtimes reported by \citet{2019MNRAS.486.2290O} range from 200 to 500 s and follow quite closely the expected barycentric correction at each midtime. Thus we hypothesize that \citet{2019MNRAS.486.2290O} used JD times in their light curves instead of either HJD/TT (in the older version of the CoRoT pipeline that they worked with) or BJD/TDB (in the Legacy version used in this work). Unfortunately, this error was then propagated to \citet{2022ApJS..259...62I}, who labeled the data as being in BJD/TDB.

\begin{figure}
    \centering
    \includegraphics[width=\textwidth]{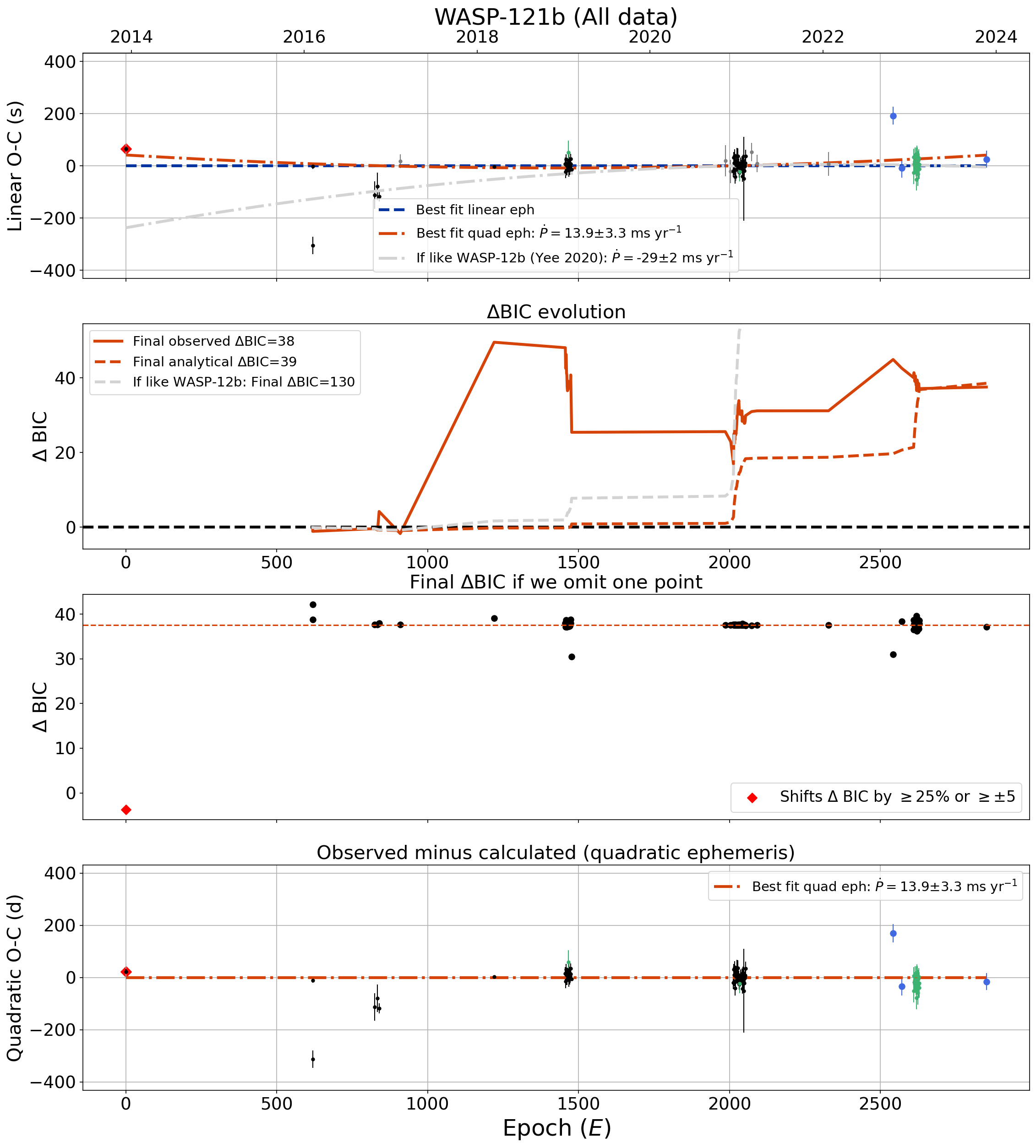}
    \caption{Timing results for WASP-121 b. Description of data same as \autoref{fig:wasp12b_timing}.}
    \label{fig:wasp121b_timing}
\end{figure}

\subsection{WASP-121 b: Possible Period Increase Heavily Dependent on Single Composite Time at Earliest Epoch}
\label{sec:wasp121}

WASP-121 b has $P=1.27$ d and was discovered by \citet{2016MNRAS.458.4025D}. We report three new light curves from 2022 to 2023 with the SAAO 40-inch (\autoref{table:obs_settings}) and use an additional 59 times from the literature and 6 times from the ETD. The first point in the sequence is a composite of nine high-quality transits taken between 9 December 2013 and 29 December 2014 \citep{2016MNRAS.458.4025D}. Note that all but five of the literature times are from two TESS sectors, making this a very sparsely observed target, with large gaps in timing coverage. For that reason, we did not remove the initial composite time, since it would have removed years of timing baseline. Fitting those times individually would provide an excellent constraint on the timing, but unfortunately these data have not been published outside of a figure in the discovery paper. 

Our timing analysis is shown in \autoref{fig:wasp121b_timing}. Not surprisingly, the value for $\Delta {\rm BIC}$ is highly dependent on the midtime from \citet{2016MNRAS.458.4025D}, as shown by the omit-one test; removing it causes the effect to entirely vanish as $\Delta {\rm BIC}$ drops below 0.  However, this system is also the only other system (besides WASP-12 b) that does not see its $\Delta {\rm BIC}$ value evaporate when the error bars are rescaled upward (see \autoref{fig:sorted_by_delta_bic_all}), retaining the second-highest rescaled $\Delta {\rm BIC}=14$ (all data, rescaled errors). This indicates that the system has reasonable error bars on most data but needs additional epochs of observations,  through both additional new light curves and a reanalysis, if possible, of the oldest data as individual transit midtimes. 

The WASP-121 b system is remarkably poorly sampled in time, with few amateur light curves in the ETD and very little published transit midtime data. Given its possible detection of period increase and its short orbital period, this planet should remain a high priority for future observations.

\subsection{WASP-46 b: Evidence for Period Increase Is Murky }
\label{sec:wasp46}

\begin{figure}
    \centering
    \includegraphics[width=\textwidth]{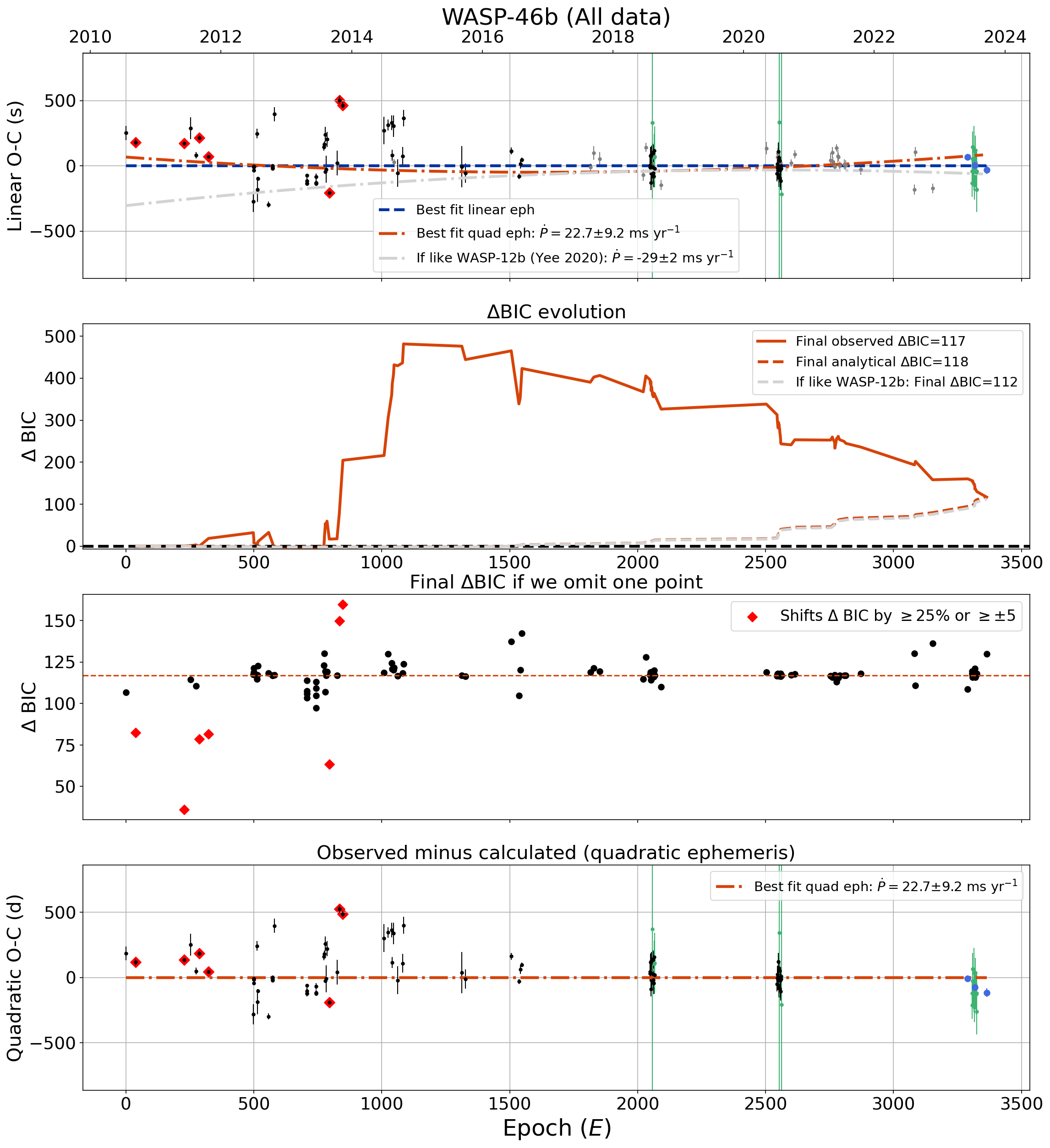}
    \caption{Timing results for WASP-46 b. Same caption as \autoref{fig:wasp12b_timing}.}
    \label{fig:wasp46b_timing}
\end{figure}

WASP-46 b, with $P=1.43$ d, was discovered by \citet{2012MNRAS.422.1988A} and is shown in \autoref{fig:wasp46b_timing}. We observed three new transits in 2023 with the SAAO 40 inch telescope (\autoref{table:obs_settings}) and fit 20 transits from the latest TESS Sector (Sector 67). In the timing analysis, we used 86 transit midtimes from \citet{2022ApJS..259...62I} excluding one composite time, which included two sectors of TESS data. We also used 28 nonduplicated light curves from the ETD with DQ=1 or 2 and errors under 5 minutes. 

Several previous timing analyses of WASP-46 b exist, including \citet{2016MNRAS.456..990C, 2017MNRAS.471..650M, 2018MNRAS.473.5126P, 2021AstL...47..638D}. Since each of these papers also included different subsets of ETD transits, care had to be taken to avoid duplication of data points. No evidence for timing change was seen in \citet{2018MNRAS.473.5126P}, though an increase in period was noted by \citet{2021AstL...47..638D}. We see a preference for a quadratic model with positive curvature, with $\Delta {\rm BIC}$=109 and a 2.6$\sigma$ detection of $\dot P = 21.6\pm8.2$ ms yr$^{-1}$. However, it is important to note that the data before about 2015 have very high scatter in the transit timing residuals, and seven points have been flagged as outliers that, if omitted individually, cause the value for $\Delta {\rm BIC}$ to range from 32 to 149 (see \autoref{fig:wasp46b_timing}). No obvious signs of timing conversion errors were found on investigating the source of those early midtimes. We also note that when the errors are rescaled, the value of $\Delta {\rm BIC}$ plummets to 2 -- still a slight preference for a nonlinear model and among the highest values reported for rescaled $\Delta {\rm BIC}$ values (see \autoref{fig:sorted_by_delta_bic_all}), but by no means a smoking gun. To resolve this murkiness, we recommend WASP-46 b for further analysis and regular transit observations.

\subsection{TrES-1 b: Period Potentially Decreasing, But Orbital Decay Unlikely}
\label{sec:tres1}

TrES-1 b, with a period of $P=3.03$ d, was one of the first transiting planets discovered \citep{2004ApJ...613L.153A} and has a very long baseline of observations, as well as prior tentative claims of period decrease \citep{2022AJ....164..220H,2022ApJS..259...62I}. Two transits of TrES-1 b were taken with MORIS on the IRTF in 2009, and we have also fit the two most recent sectors of TESS data (Sectors 53 and 54). We used 61 previously published midtimes from \citet{2022ApJS..259...62I} and another 49 midtimes from the ETD.

\begin{figure}
    \centering
    \includegraphics[width=\textwidth]{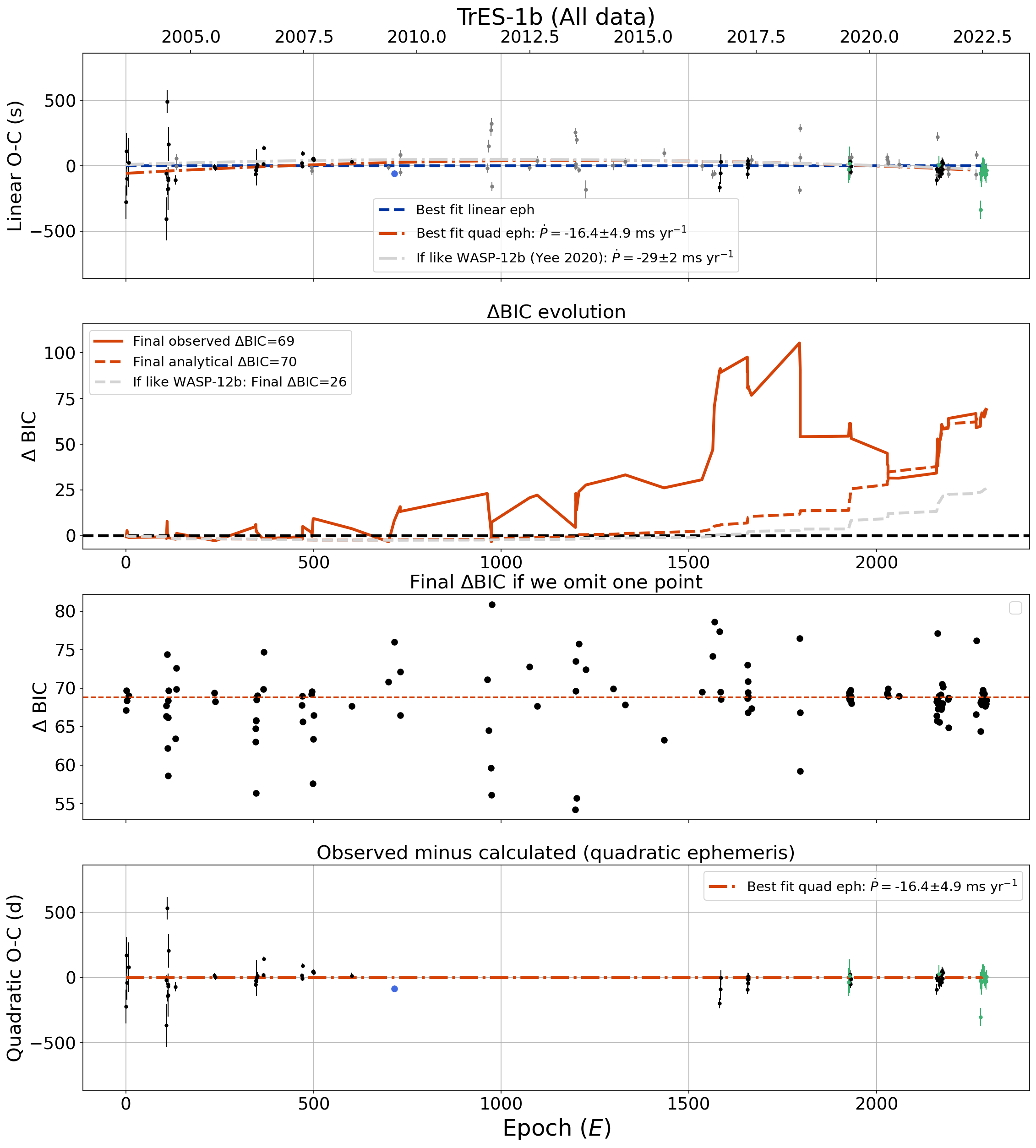}
    \caption{Timing results for TrES-1b. Description of data same as \autoref{fig:wasp12b_timing}.}
    \label{fig:tres1b_timing}
\end{figure}

In contrast with other systems, no obvious errors were identified in the literature times for TrES-1 b, and none of the midtimes were significant in our omit-one test. We found $\Delta {\rm BIC=69}$ and a 3.4$\sigma$ detection of $\dot P = -16.4\pm4.9$ ms yr$^{-1}$ (see \autoref{fig:tres1b_timing}). This is intermediate to, and consistent with, the value of $\dot P = -18.36 \pm 3.73$ ms yr$^{-1}$ that \citet{2022ApJS..259...62I} found, and the value of $\dot P = -16.0 \pm 3.7$ ms yr$^{-1}$ from \citet{2022AJ....164..220H}, and less significant than either.  However, assuming the apparent period decrease holds up to further observations, the explanation will likely be something other than orbital decay, since the inferred $Q_{\star}'=160$ is five orders of magnitude lower than the star's predicted $Q_{\star}'>10^7$  \citep{2024ApJ...960...50W}. Even the rapidly decaying Kepler-1658 b, which orbits a subgiant star, has an estimated $Q_{\star}'$ value of $10^4$, and there are no indications that the star TrES-1 has left the main sequence. 

A rapid ephemeris shift could result from perturbations from a massive companion. A  stellar companion is located 13.2 arcsec away and is 8.5 magnitudes fainter, though it is not clear if a star that distant could produce a perturbation with the magnitude and minimum oscillation period implied by this detection or if some other as-yet-undetected body would need to be responsible. Exploring the full range of dynamical scenarios that could lead to such a rapid $\dot P$ is beyond the scope of this work, but based on our analysis the signature for a nonlinear period is promising, and we recommend TrES-1 b for continued monitoring.



\subsection{Kepler-1658 b: Existing Data Have Promising $\Delta {\rm BIC}$}
\label{sec:kepler1658b}

We applied our analysis to another promising decay candidate, Kepler-1658b, a $P=3.8$ d planet that orbits an evolved host star ($2.9 R_S$, $1.5 M_S$). The orbital period decrease for this planet was measured as $\dot P = 131 \pm 22$ ms yr$^{-1}$  by \citet{Vissapragada_2022}. The evolved host star is probably key to explaining the rapid rate of decay: although \citet{2024ApJ...960...50W} found that g-mode dissipation is too weak to explain it, \citet{Vissapragada_2022} argued that inertial wave dissipation, combined with rapid stellar rotation, could.On the other hand, recent results from \citet{2024MNRAS.527.5131B} find that the time the star spends with $Q_\star^\prime < 10^4$ may be too short, potentially requiring nonsynchronous rotation if tidal dissipation is to explain the rate of orbital decrease. An alternative explanation could be tidally induced apsidal precession \citep{2024MNRAS.527.5131B}.

\begin{figure}
    \centering
    \includegraphics[width=\textwidth]{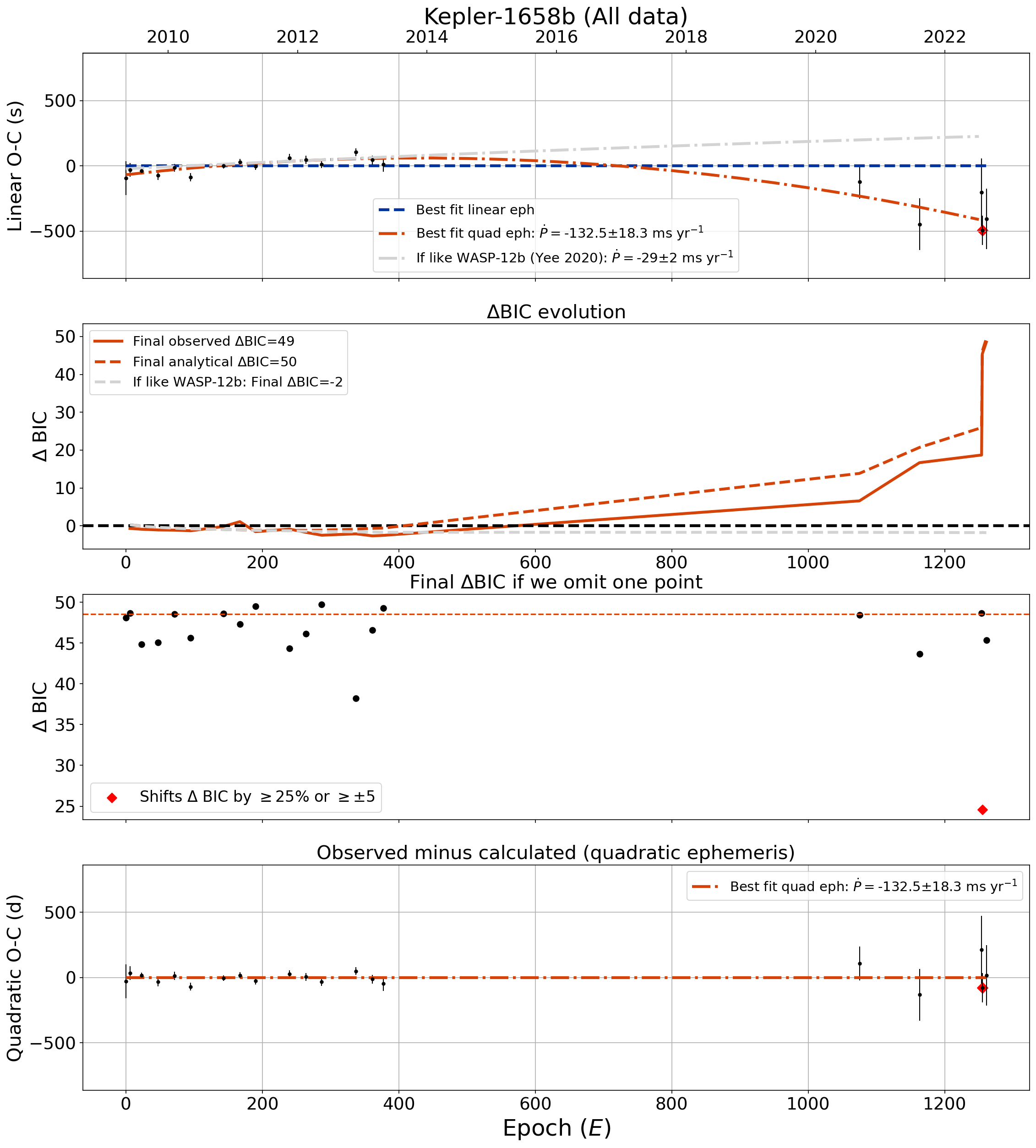}
    \caption{Timing results for Kepler-1658b, using data from \citet{Vissapragada_2022}. Description of data same as \autoref{fig:wasp12b_timing}.}
    \label{fig:kepler-1658b}
\end{figure}

Although we have no new transit observations for this planet due to its very shallow transit depth (0.1\%), we applied our analysis pipeline to the midtransit times reported in \citet{Vissapragada_2022} as described above. We found $\dot P=-132.5 \pm 18.3$ ms yr$^{-1}$, with $\Delta {\rm BIC}=49$, in agreement with \citet{Vissapragada_2022}; see \autoref{fig:kepler-1658b}. The high scatter in the photometry compared to the depth of individual transits meant that it was necessary to stack all transits in each Kepler quarter or TESS sector to produce a single time per quarter/sector. In addition, three ground-based observations were made using the Palomar 5.1-m \citep{Vissapragada_2022};  one of these points triggers our omit-one test, and without it the $\Delta {\rm BIC}$ value drops by half. Since this point has low errors and occurs near the end of the sequence, it is not surprising that it holds so much weight; this does not necessarily indicate a problem with the data but does point to the need for additional observations. We note that the nominal and rescaled values for $\Delta {\rm BIC}$ are nearly identical (49 and 50, respectively), indicating that the reported errors are likely reliably measured. The extremely rapid rate of decay for this system means that confirmation could be possible with another year or two of observations.

\begin{table}
\caption{Observational and theoretical constraints on changing period}
\small
\begin{tabular}{l | rrr rrrr | rrr } 
\hline
&\multicolumn{7}{c|}{Derived from Observations} & \multicolumn{3}{c}{Theoretical Predictions}\\
\hline
Planet &$\dot P$ & $\sigma$ & $\Delta {\rm BIC}$ & $\Delta {\rm BIC}$ & Log        &  Log              & $\tau$ & Log          & $\tau_{\rm th}$ & $|\dot P_{\rm th}|$\\
   &    &          &             & Rescaled  & $Q_\star'$  & $Q_{\star,min}$  & ~   & $Q_{\rm th}'$   &   & \\
   &  (ms yr$^{-1}$)  &          &             &   &      &   $^a$   & (Myr)   &   & (Myr)  & (ms yr$^{-1}$)\\
\hline
\multicolumn{10}{l}{\emph{Systems with $\Delta {\rm BIC}>30$}}\\
WASP-12b & $-29.8\pm1.6$ & 18.5 & 947.1 & 337.1 & $1.6$e+5 & $>$5.1 & 3 & 10 & 6e+5 & $<0.01$\\
WASP-46b & $22.6\pm9.2$ & 2.5 & 116.8 & 1.1 & -- & $>$4.8 & -- & 6 & 5e+2 & 0.16\\
TrES-1b & $-16.4\pm4.9$ & 3.4 & 68.8 & 6.4 & $1.6$e+2 & $>$1.9 & 16 & 7 & 2e+5 & $<0.01$\\
WASP-121b & $13.9\pm3.3$ & 4.2 & 37.6 & 13.0 & -- & -- & -- & 10 & 2e+6 & $<0.01$\\
\hline
\multicolumn{10}{l}{\emph{Systems with $\Delta {\rm BIC}<30$}}\\
CoRoT-2b & $-2.4\pm4.1$ & 0.6 & -1.8 & -4.7 & $8.1$e+4 & $>$4.1 & 62 & 10 & 5e+6 & $<0.01$\\
HAT-P-23b$^b$ & $-3.1\pm2.5$ & 1.3 & -0.2 & -3.2 & $3.2$e+5 & $>$5.0 & 33 & 7 & 8e+3 & 0.01\\
HAT-P-23b$^c$ & $-3.1\pm2.5$ & 1.3 & -0.2 & -3.2 & $3.2$e+5 & $>$5.0 & 33 & 10 &1e+6 & 0.01\\
HAT-P-36b & $-6.1\pm6.5$ & 0.9 & 0.0 & -4.0 & $1.3$e+5 & $>$4.5 & 19 & 6 & 3e+2 & 0.33\\
HATS-24b & $-4.5\pm14.9$ & 0.3 & -3.8 & -3.9 & $2.0$e+5 & $>$4.3 & 26 & 6 & 2e+2 & 0.42\\
HATS-35b & $-47.5\pm43.8$ & 1.1 & 0.4 & -1.9 & $5.4$e+3 & $>$3.1 & 3 & 10 & 8e+6 & $<0.01$\\
HATS-70b & $39.1\pm68.6$ & 0.6 & -2.7 & -2.7 & -- & $>$4.4 & -- & 10 & 4e+5 & $<0.01$\\
HIP-65Ab & $22.3\pm9.6$ & 2.3 & -2.1 & 0.7 & -- & $>$5.3 & -- & 5.5 & 2e+1 & 2.62\\
KELT-16b & $-20.5\pm12.5$ & 1.6 & 1.8 & -1.8 & $4.2$e+5 & $>$5.2 & 4 & 10 & 3e+5 & $<0.01$\\
KELT-1b & $7.4\pm7.8$ & 0.9 & 1.6 & -2.6 & -- & $>$6.3 & -- & 10 & 6e+4 & $<0.01$\\
KOI-13b & $-40.7\pm136.8$ & 0.3 & -1.4 & -2.0 & $5.2$e+4 & $>$3.7 & 4 & 10 & 6e+5 & $<0.01$\\
KPS-1b & $93.7\pm79.9$ & 1.2 & 2.5 & -2.6 & -- & $>$2.8 & -- & 6 & 2e+3 & 0.06\\
Qatar-10b & $-52.4\pm40.2$ & 1.3 & -1.9 & -3.7 & $3.3$e+3 & $>$3.0 & 3 & 10 & 1e+7 & $<0.01$\\
Qatar-1b & $-2.2\pm2.9$ & 0.8 & -3.5 & -5.0 & $6.6$e+4 & $>$4.1 & 56 & 6 & 6e+2 & 0.12\\
Qatar-2b & $-6.2\pm6.6$ & 0.9 & 2.7 & -3.4 & $4.6$e+4 & $>$4.0 & 19 & 6 & 4e+2 & 0.21\\
TOI-2046b & $243.7\pm107.6$ & 2.3 & 0.4 & 0.5 & -- & $>$4.0 & -- & 10 & 3e+6 & $<0.01$\\
TOI-2109b & $5.8\pm74.3$ & 0.1 & -4.5 & -4.6 & -- & $>$5.7 & -- & 8 & 2e+2 & 0.19\\
TOI-564b & $-86.0\pm36.0$ & 2.4 & -2.6 & 1.7 & $3.1$e+3 & $>$3.1 & 2 & 6 & 8e+2 & 0.12\\
TrES-2b & $-5.9\pm4.1$ & 1.4 & 4.9 & -3.4 & $3.1$e+3 & $>$3.0 & 36 & 6.5 & 8e+3 & 0.02\\
TrES-3b & $-4.0\pm1.6$ & 2.5 & 22.8 & 0.4 & $6.4$e+4 & $>$4.5 & 28 & 6 & 3e+2 & 0.27\\
WASP-18b & $-0.5\pm1.1$ & 0.5 & -4.3 & -4.5 & $3.6$e+7 & $>$6.7 & 156 & 10 & 5e+4 & $<0.01$\\
WASP-19b & $0.6\pm0.8$ & 0.7 & -4.4 & -4.9 & -- & $>$6.4 & -- & 5.5 & 2e+1 & 3.18\\
WASP-32b & $-48.7\pm27.1$ & 1.8 & 11.2 & -0.1 & $1.2$e+3 & $>$2.7 & 5 & 6.5 & 2e+4 & 0.01\\
WASP-33b & $1.2\pm4.7$ & 0.2 & -4.0 & -4.6 & -- & $>$5.1 & -- & 10 & 6e+5 & $<0.01$\\
WASP-36b & $3.7\pm3.9$ & 0.9 & -2.6 & -3.6 & -- & $>$4.5 & -- & 6 & 3e+2 & 0.28\\
WASP-3b & $-4.5\pm5.1$ & 0.9 & -1.8 & -4.2 & $7.4$e+4 & $>$4.2 & 36 & 10 & 3e+6 & $<0.01$\\
WASP-43b & $-1.0\pm1.1$ & 0.9 & -0.4 & -4.6 & $1.7$e+6 & $>$5.6 & 70 & 5.5 & 3e+1 & 1.56\\
WASP-50b & $2.6\pm12.8$ & 0.2 & -3.7 & -3.9 & -- & $>$3.1 & -- & 6.5 & 6e+3 & 0.02\\
WASP-52b & $-12.2\pm6.4$ & 1.9 & 19.3 & -1.4 & $1.4$e+3 & $>$2.7 & 12 & 6.5 & 2e+4 & 0.01\\
WASP-5b & $-4.2\pm4.0$ & 1.1 & -0.5 & -3.2 & $6.9$e+4 & $>$4.3 & 34 & 6 & 8e+2 & 0.12\\
WASP-64b & $2.3\pm6.5$ & 0.4 & -4.2 & -4.3 & -- & $>$4.1 & -- & 6 & 8e+2 & 0.12\\
WASP-77Ab & $4.1\pm7.4$ & 0.6 & -3.4 & -3.6 & -- & $>$4.3 & -- & 6 & 4e+2 & 0.20\\
WASP-103b & $-4.1\pm8.4$ & 0.5 & -3.2 & -3.8 & $1.7$e+6 & $>$5.4 & 19 & 6 & 4e+1 & 1.29\\
WASP-104b & $5.0\pm4.4$ & 1.2 & -2.7 & -3.5 & -- & $>$4.0 & -- & 6 & 1e+3 & 0.10\\
WASP-114b & $-28.4\pm76.1$ & 0.4 & -1.2 & -2.6 & $2.7$e+4 & $>$3.5 & 5 & 7 & 3e+3 & 0.03\\
WASP-135b & $-3.1\pm8.0$ & 0.4 & -3.9 & -3.8 & $1.1$e+5 & $>$4.1 & 39 & 6 & 3e+2 & 0.26\\
WASP-145Ab & $11.1\pm10.8$ & 1.0 & -2.0 & -2.0 & -- & $>$3.0 & -- & 7 & 3e+4 & $<0.01$\\
WASP-163b & $-8.2\pm132.0$ & 0.1 & -2.7 & -2.7 & $3.3$e+4 & $>$2.8 & 17 & 6 & 6e+2 & 0.15\\
WASP-164b & $-54.5\pm25.1$ & 2.2 & 13.6 & 1.2 & $2.5$e+3 & $>$3.0 & 3 & 6 & 8e+2 & 0.12\\
WASP-173Ab & $19.3\pm11.0$ & 1.8 & 1.1 & -0.6 & -- & $>$5.0 & -- & 6 & 1e+2 & 0.60\\
\hline
\multicolumn{11}{l}{\footnotesize{a. 3$\sigma$ lower limit on  $Q_\star'$ (in log base 10) as calculated from the lower limit on $\dot P$. No value if $\dot P - 3 \sigma > 0$. }} \\
\multicolumn{11}{l}{\footnotesize{b.  HAT-P-23 theoretical calculations if using $M_\star=0.58~ M_{sun}$, $M_p=1.34~ M_J$ and $R_\star=0.58~ R_{\sun}$.  }} \\
\multicolumn{11}{l}{\footnotesize{c.  HAT-P-23 theoretical calculations if using $M_\star=1.13~ M_{sun}$, $M_p=1.97~ M_J$ and $R_\star=1.19~ R_{\sun}$. }}
\end{tabular}
\label{table:decay_params}
\end{table}
%


\subsection{Data for Marginal Systems ($\Delta {\rm BIC}<30$)}
\label{sec:marginal}

We chose $\Delta {\rm BIC}>30$ somewhat arbitrarily to select a handful of the most promising candidates to focus on in this paper. It is also a useful discriminant of where $\Delta {\rm BIC}$ values as calculated with real data and error bars start to become robust enough to have more confidence that they are measuring something other than noise. Many of the systems with possible claimed detections listed in the Introduction were presented at much smaller $\Delta {\rm BIC}$ values.  We will now briefly discuss our analysis of the 37 planets that fall below $\Delta {\rm BIC}<30$ that have not been described above, as shown in \autoref{fig:sorted_by_delta_bic_all}.  Our preliminary values for $\Delta {\rm BIC}<30$ and the best-fit quadratic model parameters are listed in \autoref{table:decay_params}. We add an important caveat that unlike with the 6 main targets of this paper, for these 37 planets we have \emph{not} done a deep dive into the sources of the literature times used. As we have demonstrated for other systems above, it is quite possible that there may be additional problems lurking in the data for one or more of these systems. 

For each of these 37 planets, we used literature times from \citet{2022ApJS..259...62I} and the ETD. We removed duplicate entries and have limited the ETD data to DQ=1 or 2 for most targets. (For WASP-163 b, which has very few ETD points, we allowed all DQ=1-5, while for Qatar-1 b, TrES-3 b, and WASP-43 b, which have many transits, we used only DQ=1.) We have not searched for additional transit midtimes that were not in \citet{2022ApJS..259...62I}, including systems for which there is known Kepler or K2 data. We have, however, analyzed additional sectors of TESS data where available. For systems that are recently discovered or have very few published literature times, we refit some literature light curves that had not previously had individual fits, using data provided in private correspondence with the authors of those works. These systems are: HATS-70 b \citep[6 light curves,][]{2019AJ....157...31Z}; NGTS-1 b \citep[1 light curve and 1 NGTS composite light curve,][]{2018MNRAS.475.4467B}; Qatar-10 b \citet[7 light curves,][]{2019AJ....157..224A}; WASP-43 b \citet[3 light curves,][]{2022MNRAS.516.4684F}. These analyses will be part of forthcoming publications similar to this one, and the values presented in \autoref{table:decay_params} should be considered preliminary results.

For all systems with nominal $\Delta {\rm BIC}<30$, few retained a positive value for $\Delta {\rm BIC}$ when we rescaled  the error bars to determine the impact of underestimated errors on $\Delta {\rm BIC}$ (see discussion of test in Section \ref{sec:calc_eph}). Of the four high-$\Delta {\rm BIC}$ ($\Delta {\rm BIC}>30$) planets, all still have positive rescaled $\Delta {\rm BIC}$ values, as noted above. But among the four intermediate planets ($10<\Delta {\rm BIC}<30$), only two, WASP-164 b and TrES-3 b, still have positive rescaled $\Delta {\rm BIC}$ values, and among the 8 planets with slightly positive nominal values ($0< \Delta {\rm BIC}<10$), just one, TOI-2046 b, is still positive in the rescaled test. Two other planets, TOI-564 b and HIP 65A b, had small negative nominal $\Delta {\rm BIC}$ values that switched to small positive rescaled $\Delta {\rm BIC}$ values. The last three planets are all sparsely sampled with large errors on individual transits and only 4-5 years of baseline, precisely the kinds of systems where swings in $\Delta {\rm BIC}$ are common \citep{2023AJ....166..142J}, and more data are needed before the true patterns of these systems will become clear. We interpret the shifts in $\Delta {\rm BIC}$ to be suggestive that underestimated errors alone may explain most of the marginal detections of decreasing orbital period, both here and in the literature.

Of the four planets in \autoref{fig:sorted_by_delta_bic_all} with intermediate values of $10<\Delta {\rm BIC}<30$, TrES-3 b, WASP-52 b, WASP-164 b, and WASP-32 b, all have decreasing periods, though none have significant values of $\dot P$ ($1.6<\sigma \le 2.5$). Of these, two have previous claimed detections: (1) A possible decreasing period for TrES-3 b has been noted at similar marginal levels by other works \citep{2022AJ....164..220H,2022AJ....164..198M}. Given that TrES-3 b is a long-studied planet around a bright star with particularly deep transits, hundreds of diverse light curves have been recorded by amateurs and professionals alike (with over 600 light curves listed in the ETD alone), making the chance of there being undetected mistakes in the literature high. A thorough reanalysis of the available record needs to be undertaken before any claims for or against period change could be credibly believed for TrES-3 b. (2) \citet{2023MNRAS.520.1642S} noted possible nonlinearity for WASP-32 b, though that work preferred an apsidal precession model to the quadratic orbital decay model. Neither their value for $\dot P$ nor ours is significant (both around 2 $\sigma$), and the system is relatively sparsely observed. Additional observations are required to assess this claim.


\section{Preliminary Population Analysis: Where Are the Decaying UHJs?}
\label{sec:population_and_qstar}

\subsection{Observational Constraints: In What Systems Could We Even Detect WASP-12 b-Like Decay?}

Most UHJs in our sample prefer a linear ephemeris (27 out of 43, or 63\%), and of those that do not, only 10 show a preference for a decreasing orbital period model (see \autoref{fig:sorted_by_delta_bic_all}). Even fewer UHJs show any preference for a decreasing orbital period when we rescale $\Delta {\rm BIC}$ to account for underestimated error (5 out of 43, or 12\%), and the dropoff between ranked values is stark: while WASP-12 b boasts a rescaled $\Delta {\rm BIC}>300$, the second-highest rescaled $\Delta {\rm BIC}$ is for TrES-1 b, at 6.4, and the remaining three planets all fall below 2, indicating very marginal model preference. This lack of detection of period change for most systems presents an interesting question that we would like to take a first pass at answering: where \emph{are} the decaying UHJs? Put another way: was WASP-12 b the first doomed world to be detected because it was one of the first to have enough transits and a long enough observational time span, or is there some feature of the WASP-12 system (e.g., its potential status as an evolved star) that sets it apart from most UHJ systems?

As noted in the Introduction, no fewer than twelve planets, not counting WASP-12 b and Kepler-1658b, have had cases presented in the recent literature that performed similar analyses and presented weak hints of period change. While not all of those planets have yet been included in our observing program, it is interesting to note that of the four that we have observed, we find instead a strong preference for a linear ephemeris for two (WASP-19 b using corrected data, and WASP-43 b) and insignificant evidence for nonlinearity for the other two (TrES-2 b and TrES-3 b). Meanwhile, our single plausible candidate for a decreasing orbital period, TrES-1 b, is decreasing too quickly to be likely to be due to orbital decay (see Section \ref{sec:tres1}). 

Should we expect to have detected orbital decay in a substantial number of systems? To answer this question, we performed two simple tests to determine if we could have even detected decay as rapid as that of WASP-12 b. Theoretically, we would expect a wide range of decay rates commensurate with the diverse stellar and planetary properties for these systems (see the range of predicted values in \autoref{table:decay_params}), but using WASP-12 b's relatively rapid rate sets a useful baseline against which to test our ability to detect tidal decay. The results of these tests are listed in \autoref{table:notlikewasp12}. The first test assigns all planets the same rate of decay as WASP-12 b, $\dot P = -29$ ms yr $^{-1}$ \citep{2020ApJ...888L...5Y}.  We find that nearly half (18 out of 42 planets that are not WASP-12 b) would have had $\Delta {\rm BIC}>30$, as shown in the Test 1 column in \autoref{table:notlikewasp12}, while in reality only one did -- and that planet, WASP-121 b, has a \emph{positive} value for $\dot P$. Four systems (WASP-19 b, WASP-43 b, WASP-18 b and TrES-3 b) would have $\Delta {\rm BIC}$ values up to ten times larger than that of WASP-12 b if they were decaying at the same rate, making the continued non-detection of decay in these systems highly significant.

We framed the second test to see what would have been observed if all stars had the same tidal dissipation factor. We assumed that every star had the same low value of $Q_\star' = 1.8 \times 10^5$ as WASP-12 b, also from \citet{2020ApJ...888L...5Y}. We then calculated the rate of change $\dot P$ that would imply, and if the calculated values is equal to or greater than 3 times the observed error then the planet is listed in \autoref{table:notlikewasp12}. Seven planets, five of which also passed Test 1, would have had $3\sigma$ results for $\dot P$ under Test 2 (\autoref{table:notlikewasp12}), including WASP-19 b, WASP-43 b, and WASP-18 b. Interestingly, the one planet for which we have the strongest claim for negative $\dot P$, TrES-1 b, did not pass this test: if it had a WASP-12 b like $Q_{\star}'$ we would not have detected any change in period.

Thus, for 20 out of 42 systems, there is evidence that we could have detected WASP-12 b like decay, but have not, meaning these planets are likely to not be decaying as rapidly. The flip side of this statement is that for the other half of our sample (22/42 planets), we still do not have enough baseline to even detect decay as rapid as WASP-12 b. With the shortest of these planets having baselines that extend just 4 years, it will take several more years to perhaps even decades to determine just how unusual WASP-12 b is. 

\subsection{Observational Constraints: Lower Limits on $Q_\star'$}

In \autoref{table:decay_params} we report values for $Q_\star'$ for systems with negative values of $\dot P$, even if they are statistically consistent with zero tidal decay. Recall that the value of $Q_\star'$ is inversely proportional to the rate of period change; the approximation used here is based on the formulation of \citet{1966Icar....5..375G}:


\begin{equation}
    Q_\star' = -\left( \frac{27\pi}{2} \right) \left(\frac{dP}{dt}\right)^{-1} \left(\frac{M_p}{M_s} \right) \left(\frac{a}{R_\star}\right)^{-5}.
\end{equation}

For every system, we used the 3$\sigma$ lower error to determine what the maximum decay rate is that would be consistent with the errors on $\dot P$, and if so what the lower limit on $Q_\star'$ must be for it to not have been detected by our survey. For systems with a nominal negative value for $\dot P$, we also calculated an estimated decay lifetime, $\tau$. For WASP-121 b, $\dot P$ is more than 3$\sigma$ above zero and we cannot place any constraint on $Q_\star'$, even a lower limit. For most of the systems the constraints placed on $Q_\star'$ are much lower than any theory would predict, but we find that nine systems (9 out of 41, or 22\%) must have $\log_{10} Q_\star' \ge 5.1$, the 3$\sigma$ lower limit for WASP-12 (\autoref{table:decay_params}). Three systems (7\%) can constrain $\log_{10} Q_\star'$ to be at least an order of magnitude larger than that of the star WASP-12: WASP-18 (6.7), WASP-19 (6.4), and KELT-1 (6.3). Again, it is important to note that we have neglected contributions to orbital decay from tidal dissipation within the planets. Therefore, the constraints here on $Q_\star^\prime$ are lower limits in this sense, too. Contributions from dissipation within the planets would allow $Q_\star^\prime$ to be even larger than we report here and still be consistent with our observations.

\begin{figure}
    \centering
    \includegraphics[width=\textwidth]{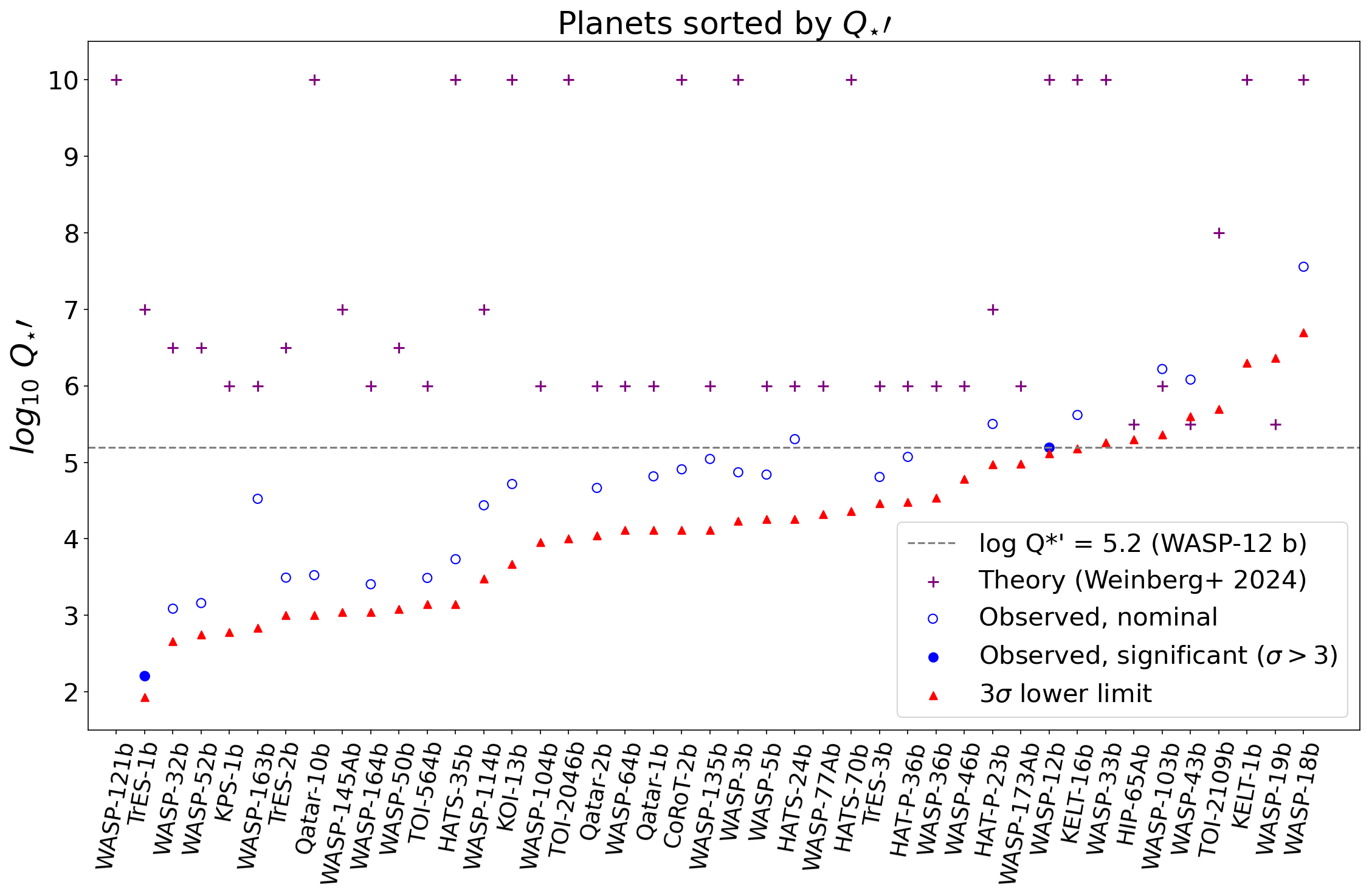}
    \caption{Constraints on $Q_\star'$ placed by observations. Lower limits are shown as upward facing triangles, and represent the 3$\sigma$ lower limit for $Q_\star'$; these were calculated even for systems that had marginally positive values for $\dot P$ (increasing periods), so long as the 3$\sigma$ value is below zero (only WASP-121 b does not have this constraint). If $\dot P$ is nominally negative but insignificant ($<3\sigma$ detection), the corresponding value for $Q_\star'$ is shown as an open blue; the two systems with significant values for $\dot P$, WASP-12 b and TrES-1 b, are shown as filled blue circles. Finally, we plotted the theoretical estimates for $Q_\star'$ from \autoref{table:decay_params} based on the work of \citet{2024ApJ...960...50W} as purple plus signs. See \autoref{table:decay_params} for all parameter values.} 
    \label{fig:qstar_constraints}
\end{figure}

\subsection{Comparison to theory}

To place our observational constraints in a theoretical context, we used the grids from \citet{2024ApJ...960...50W} to estimate $Q_\star'$ for each system. We used Figures 5 and 7 from that paper and the best-fit system parameters from \autoref{table:system_params}. Because the grid of calculated $Q_\star'$-values from that study is very coarse, instead of formally interpolating, we used a by-eye approximation. A major source of uncertainty is the large range in stellar age estimates, so we used the central value of the age estimate (when it exists). Note that HAT-P-23 shows up twice in \autoref{table:decay_params} because of conflicting stellar mass estimates in the NASA Exoplanet Archive; we include both to illustrate how stellar uncertainty impacts these calculations.

The resulting theoretical estimates for $Q_\star'$ are listed as $Q_{\rm th}'$ in \autoref{table:system_params}, and we computed the theoretical decay time $\tau_{\rm th}$ and rate of period decrease $\dot P_{\rm th}$ as in Eq. 15 in \citet{2024ApJ...960...50W}. For stellar models with radiative cores and convective envelopes, \citet{2024ApJ...960...50W} found values that range from $Q_{\rm th}' \approx \sim10^5$ (at $P=0.5 \textrm{ d}$) to $Q_{\rm th}' \approx 10^6-10^7$ at $P=2 \textrm{ d}$. For higher-mass stars ($>1.2 M_\sun$) on the main sequence, the cores are convective and in such cases $Q_\star'$ is likely to be very large; for all such stars, we set  $Q_{\rm th}'=10^{10}$ as a representative value. If those stars were instead subgiants, however, as might be the case for WASP-12 b, their $Q_\star'$ would be much smaller since their cores would become radiative.
 
Four systems are predicted, based solely on these theoretical estimates, to have rapid orbital decay as a result of having predicted $\log_{10} Q_\star' = 5.5-6$, with a theoretical prediction of $\dot P_{\rm th} < -1.0 \textrm{ ms yr}^{-1}$ for HIP 65 Ab, WASP-103 b, WASP-19 b, and WASP-43 b (see \autoref{table:decay_params}). In two cases, WASP-19 b and WASP-43 b, the theoretical estimates (both have $\log_{10} Q_{\rm th}' = 5.5$) are actually lower than the current limits placed by our observations (respectively $\log_{10} Q_{\rm th}'>6.4$ and $\log_{10} Q_{\rm th}'>5.6$), as can be seen in \autoref{fig:qstar_constraints}.  However, in both cases the stellar age uncertainties are large: the age of WASP-19 is $5.5^{+8.5}_{-4.5} \textrm{ Gyr}$ while WASP-43's age is $7.0 \pm 7.0 \textrm{ Gyr}$. If the stars were closer to the younger age, then we would predict correspondingly higher values for $\log_{10} Q_{\star}'$ and significantly slower decay rates (see panels on left side of Fig 5 in \citealt{2024ApJ...960...50W}).

\begin{table}
\caption{Could Have Detected WASP-12 b-Like Decay But Did Not}
\small
\begin{tabular}{l | r | r || r | r} 
\hline
     	 & Observed   	&  Test 1 	 & Observed 	 & Test 2 \\
Planet    	 & $\Delta$BIC 	   	& Calc. $\Delta$BIC  	 & $\dot P$ 	 & Calc. $\dot P$  \\
        	&    ~  	 & if $\dot P$ like W12$^a$  	 &  ~  	 & if $Q_{\star}'$ like W12$^b$  \\
\hline
WASP-12b	 &947.1	 &896.1	 &$-29.8\pm1.6$	 &$-27\pm1$\\
\hline
WASP-43b	 &-0.4	 &7443.7	 &$-1.0\pm1.1$	 &$-10\pm1.1$\\
WASP-19b	 &-4.4	 &5681.1	 &$0.6\pm0.8$$^c$	 &$-23\pm0.8$\\
WASP-18b	 &-4.3	 &1597.9	 &$-0.5\pm1.1$	 &$-110\pm1.1$\\
TrES-3b	     &22.8	 &1041.9	 &$-4.0\pm1.6$	 &--\\
WASP-33b	 &-4.0	 &335.0	 &$1.2\pm4.7$$^c$	 &--\\
HAT-P-23b	 &-0.2	 &314.3	 &$-3.2\pm2.5$	 &--\\
Qatar-1b	 &-3.5	 &208.5	 &$-2.2\pm2.9$	 &--\\
CoRoT-2b	 &-1.8	 &180.4	 &$-2.4\pm4.1$	 &--\\
WASP-121b	 &37.6	 &129.8	 &$14.0\pm3.3$$^c$	 &--\\
Qatar-2b	 &2.7	 &98.5	 &$-6.2\pm6.6$	 &--\\
WASP-5b	     &-0.5	 &77.9	 &$-4.2\pm4.0$	 &--\\
HAT-P-36b	 &0.0	 &70.2	 &$-6.1\pm6.5$	 &--\\
KELT-1b	    &1.6	 &60.8	 &$7.4\pm7.8$$^c$	 &$-186\pm7.8$\\
WASP-103b	 &-3.2	 &56.1	 &$-4.1\pm8.4$	 &$-40\pm8.4$\\
WASP-36b	 &-2.6	 &55.6	 &$3.7\pm3.9$$^c$	 &--\\
WASP-52b	 &19.3	 &49.5	 &$-12.2\pm6.4$	 &--\\
TrES-2b	     &4.9	 &44.1	 &$-5.9\pm4.1$	 &--\\
WASP-3b	     &-1.8	 &42.1	 &$-4.5\pm5.1$	 &--\\
KELT-16b	 &1.8	 &--	 &$-20.5\pm12.5$	 &$-50\pm12.5$\\
TOI-2109b	 &-4.5	 &--	 &$5.8\pm74.3$$^c$	 &$-641\pm74.3$\\
\hline
\multicolumn{5}{l}{\footnotesize{a. Test 1: analytical $\Delta$BIC $>30$ if WASP-12 b like decay ($\dot P= -29$ ms y$^{-1}$) \citep{2020ApJ...888L...5Y} }}\\
\multicolumn{5}{l}{\footnotesize{c. Test 2: $\dot P$ significant by $\ge3\sigma$ if $Q_\star' = 1.8 \times 10^5$ \citep{2020ApJ...888L...5Y} }}\\
\multicolumn{5}{l}{\footnotesize{c. Note: measured $\dot P$ is actually positive (increasing period). }}
\end{tabular}
\label{table:notlikewasp12}
\end{table}

\section{Discussion and Conclusions}
\label{sec:disc}

\subsection{No Evidence Yet for Orbital Decay Being a Common UHJ Trait}

Of the 43 planets with updated timing information presented in this work, just one -- WASP-12 b -- has clear indications of period decrease that is likely due to orbital decay. In this work, we identify another system, TrES-1 b, which shows signs of a decreasing orbital period, but at too rapid a rate to be due to orbital decay under any known theoretical prediction. We also have identified two systems, WASP-121 b and WASP-46 b, which show signs of increasing orbital period, though both systems have issues with existing data that may explain some or all of their purported period change. 

For nearly half of our targets that are not WASP-12 b (20 out of 42), a case can be made that they are not experiencing orbital decay as rapidly as WASP-12 b, or else we would have detected a $3\sigma$ value for $\dot P$ or $\Delta {\rm BIC}>30$, or both. For a few systems (notably WASP-19 b, KELT-1 b, and WASP-18 b) the observations constrain orbital decay to be at least an order of magnitude slower, and $Q_{\star}'$ at least an order of magnitude greater, than that of the WASP-12 b system.

We note that of the two systems for which orbital decay is a credible explanation, both Kepler-1658 b \citep{Vissapragada_2022} and WASP-12 b \citep{2017ApJ...849L..11W} may have evolved off the main sequence, which could offer an explanation for more rapid decay \citep[e.g.,][]{2024ApJ...960...50W}. However, the status of the star WASP-12 is ambiguous \citep{2024A&A...686A..84L}, with a main sequence star still favored \citep{2019MNRAS.482.1872B}. If WASP-12 is in fact evolved, that would leave no main sequence stars known to host decaying planets, offering support to predictions that orbital decay is unlikely during the main sequence and is a feature of later stages of stellar life. This is however in tension with the independent observation that hot Jupiter host stars are younger than all planet hosts \citep{2019AJ....158..190H, 2023PNAS..12004179C, 2023AJ....166..209M} and that tidal decay does occur on the main sequence \citep{2019AJ....158..190H}. It is thus important to search more UHJs both on and off the main sequence for doomed worlds to see if this suggestive connection holds beyond a sample size of two planets.

We also note that for at least half of our sample, we lack sufficient data to have detected orbital decay even as rapid as that of WASP-12 b. It may be the case that identification of decaying planets around stars on the main sequence will require additional years of observations from either space (e.g., TESS) or ground-based observatories, in order for slower rates of orbital decay to reveal themselves.  Based on calculations in this paper we find that values for $Q_\star'$ probably span at least three orders of magnitude ($\sim10^4$ for Kepler-1658b, $\sim10^5$ for WASP-12 b, and $>\sim10^7$ for WASP-18b). Predictions for the rates at which planets undergo orbital decay will have to take into account stellar mass and age.

\subsection{Best Practices for Long-term Timing Analyses}

We present three recommendations about best practices for timing analyses to search for orbital decay: 

\begin{enumerate}

    \item \textbf{$\Delta {\rm BIC}$ is a powerful tool for identifying systems of interest, especially accompanied by a significant value for $\dot P$. However, it is only as good as the error bars of the accompanying data.} With the exception of a uniform dataset for which the errors are well known (e.g., stacking all of the transits in a single Kepler quarter or TESS sector, as was done for Kepler-1658b in \autoref{fig:kepler-1658b}), it is likely that some errors are underestimated, increasing the apparent values for $\Delta {\rm BIC}$. Thus, when using $\Delta {\rm BIC}$ with real data, it is important to understand that light curves in the published literature may have mistakes in both the absolute timing and in the size of the nominal error bars. We recommend care in interpreting $\Delta {\rm BIC}$ results and suggest the tests described in this work, including the omit-one test to identify points that may be having an out-sized impact on the apparent result, and the rescaled $\Delta {\rm BIC}$ test to identify if overall errors on the system may be too small. Given our experience in this work $\Delta {\rm BIC}$ values below 30-50 should be treated as tentative detections in need of confirmation.

    \item \textbf{In the ideal case, someone would observe a transit of every planet that might be experiencing orbital decay at least once a year to avoid long gaps in the timing data}, following the recommendation of \citet{2023AJ....166..142J}. We note for instance that WASP-121 b, which has very few distinct epochs available, would be a much clearer case if it had been more regularly observed in the years following its first announcement. Even relatively modest aperture telescopes, such as the 14-inch used in this work, are useful since many UHJs are found around sufficiently bright stars to constrain the transit timing. Widespread networks of small scopes may also provide robust facilities for keeping ephemerides fresh and improving prospects for tidal-decay detection \citep{2023PASP..135a5001P}. In fact, phasing observations of a single transit conducted by multiple small telescopes simultaneously could result in a very high effective timing resolution, with the potential to significantly reduce mid-transit time uncertainties.

    \item \textbf{There are timing errors in the literature, and the current system tends to propagate rather than eliminate errors.} We identified errors spanning multiple published works in the timing of CoRoT-2 b and WASP-19 b. These errors had a strong effect, in that correcting them removed the signal of tidal decay entirely; but other errors may well be lurking that conform more to our expectations and thus have not been brought to light. We discuss this last item in detail in the next section.

\end{enumerate}

\subsection{The Timing Community Has a Database Problem}
\label{sec:database}

It is becoming clear that the scientific record is currently not well set up to facilitate long-term timing searches for slow changes over time involving multiple instruments and observers over many years. Timing errors, particular conversion between times in UTC (what most data are recorded in) and BJD/TDB (what we need to use to detect true dynamical changes), are easy to make, hard to detect, and even harder to correct for posterity. Additionally, far too many early transit light curves have never been published except as figures in papers, removing valuable years of observational baseline from reanalysis. In some cases, not even a fitted midtime is available for individual light curves. And the current model of publication, whereby each new analysis publishes a list that combines midtimes of newly observed transits and times taken from the literature, has led to compilations of compilations of midtimes. We have documented multiple cases where this process has introduced errors in one work that are then propagated through multiple works.

The problem of data curation currently constitutes a major component of all long-term timing analyses. Rather than each group repeating the same effort to identify, extract, convert, reformat, and refit prior transit light-curve data, there should be a single archival database that contains a definitive list of all prior published transit times, as taken directly from their original publication. If more than one work has reported a time for a given transit, it should be clear which times are re-analyses, and a best time should be indicated, similar to how the NASA Exoplanet Archive reports all values for stellar and planetary parameters but also provides a best value for researchers to use. This ideal transit-timing database would provide information about each individual light curve, and make transit light curves available for download in a uniform format to make it easy to re-fit data that is supposed to be in the public domain, but in practice has been inconvenient to access. The Exoplanet Transit Database is an excellent resource and is well used by amateurs, but it is missing some key features (notably, it uses HJD/UTC instead of BJD/TDB timing and lacks a bulk download option for both light curves and the original JD times before conversion) and, moreover, has low uptake from the professional community. It would require either a substantial overhaul of the ETD or an entirely new system to be developed and funded for the long term. 

In addition to observations of primary transits, an ideal timing database would also include timing constraints placed by planetary occultations (also called secondary eclipses). Occultations are even less likely to have published timing data available for individual light curves, owing to both their lower signal-to-noise and to the propensity of occultation data to be used to study planetary atmospheres, where stacked and phased data of multiple occultations are necessary to tease out small signals. Occultation timing observations are used to determine if apsidal precession could explain a timing trend, since orbital decay will cause the interval between both transits and occultations to decrease, while precessing systems would have opposite trends, with occultations showing increasing periods while transits show decreasing periods, and vice versa.

Additionally, there should be an expectation shift that publishing data for a transit or occultation light curve requires publishing the photometric time series. Transit light curves can be archived as text files and do not require much data storage space compared to many astronomical datasets. Most journals can easily accommodate this dataset as supplementary data files. Ideally, the published photometric light curves would also be required to be archived in a common format in the uniform transit timing database.

Finally, a thorough archival effort should be made to retrieve transit and occultation light curves that were used in prior works and may still exist on the hard drives of the original researchers, but which have not been made public, before the data are completely lost. These early transits and occultations provide critical data for constraining orbital decay. Without them, we will just have to wait longer to identify which worlds are doomed.

\section*{Acknowledgments}
This study was supported by a grant from NASA's Exoplanet Research Program 21-XRP21-0170.

This paper uses observations made at the South African Astronomical Observatory (SAAO).

E.R.A. and A.A.S. were visiting astronomers at the Infrared Telescope Facility, which is operated by the University of Hawaii under contract 80HQTR19D0030 with the National Aeronautics and Space Administration. The authors wish to recognize and acknowledge the very significant cultural role and reverence that the summit of Maunakea has always had within the indigenous Hawaiian community. We are most fortunate to have the opportunity to conduct observations from this mountain, and we recognize that we are guests.

Data presented herein were obtained at the \textsc{Minerva}-Australis facility from telescope time allocated under the NN-EXPLORE program with support from the National Aeronautics and Space Administration.

\textsc{Minerva}-Australis is supported by Australian Research Council LIEF Grant LE160100001, Discovery Grants DP180100972 and DP220100365, Mount Cuba Astronomical Foundation, and institutional partners University of Southern Queensland, UNSW Sydney, MIT, Nanjing University, George Mason University, University of Louisville, University of California Riverside, University of Florida, and The University of Texas at Austin.

We respectfully acknowledge the traditional custodians of all lands throughout Australia and recognise their continued cultural and spiritual connection to the land, waterways, cosmos, and community. We pay our deepest respects to all Elders, ancestors and descendants of the Giabal, Jarowair, and Kambuwal nations, upon whose lands the \textsc{Minerva}-Australis facility at Mt. Kent is situated.


%






\bibliography{main}{}
\bibliographystyle{aasjournal}



\end{document}